\documentclass{eptcs}
\providecommand{\event}{VPT 2017}
\usepackage{breakurl}
\usepackage{underscore}

\usepackage[table,xcdraw]{xcolor}
\usepackage{algorithm}
\usepackage{algpseudocode}
\usepackage{amsmath} % needed for the split environment
\usepackage{amsfonts}
\usepackage{amssymb}
\usepackage{amsthm}
\usepackage{caption}
\usepackage[draft]{fixme}
\fxsetup{theme=color}
\usepackage{fancyvrb} % for two-column code layout
\usepackage{float} % for mylisting float
\newfloat{mylisting}{h}{mylistings}
\floatname{mylisting}{Listing}
\usepackage{graphicx}
\usepackage{makeidx} % allows for index generation
\usepackage{placeins}
\usepackage[mode=tex]{standalone} % so standalone trees can be included
\usepackage{subcaption}
\usepackage{tikz}
\usepackage{tikz-qtree}
\usepackage{verbatimbox}
\newtheorem{definition}{Definition}

\usepackage[utf8]{inputenc}
\usepackage{amssymb}
\usepackage{tikz}
\usepackage{tikz-qtree}
\usepackage{xr}
\usetikzlibrary{decorations,arrows,shapes}
\usepackage{amsmath}

\newcommand{\importantterm}[1]{\emph{#1}}

\begin{document}
\title{Transforming Coroutining Logic Programs into Equivalent CHR Programs}
\author{Vincent Nys\thanks{Vincent Nys is supported by FWO Flanders under research contract G088414N.}
\institute{Department of Computer Science\\
KU Leuven\\
Leuven, Belgium}
\email{vincent.nys@kuleuven.be}
\and
Danny De Schreye
\institute{Department of Computer Science\\
KU Leuven\\
Leuven, Belgium}
\email{danny.deschreye@kuleuven.be}
}
\def\titlerunning{Transforming Coroutining Logic Programs into Equivalent CHR Programs}
\def\authorrunning{Vincent Nys \& Danny De Schreye}

\maketitle

\begin{abstract}
We extend a technique called Compiling Control. The technique transforms coroutining logic programs into logic programs that, when executed under the standard left-to-right selection rule (and not using any delay features) have the same computational behavior as the coroutining program. In recent work, we revised Compiling Control and reformulated it as an instance of Abstract Conjunctive Partial Deduction. This work was mostly focused on the program analysis performed in Compiling Control. In the current paper, we focus on the synthesis of the transformed program. Instead of synthesizing a new logic program, we synthesize a CHR(Prolog) program which mimics the coroutining program. The synthesis to CHR yields programs containing only simplification rules, which are particularly amenable to certain static analysis techniques. The programs are also more concise and readable and can be ported to CHR implementations embedded in other languages than Prolog.
\end{abstract}

\section{Introduction}
Compiling Control (CC for short) is a program transformation technique proposed 30 years ago in \cite{bruynooghe1989compiling} and \cite{deschreyetransformation1989}.
The technique aims to compile the runtime behavior of pure logic programs executed under a non-standard computation rule to equivalent logic programs that perform the same computations under the standard, left-to-right computation rule of Prolog.
The motivation for the technique was efficiency.

The technique is designed to work in two phases.
In a first phase, the computation flow of the program, executed under the non-standard rule, is analyzed, resulting in a symbolic evaluation tree that captures the entire flow.
In a second phase, from the symbolic evaluation tree, a new logic program is synthesized, that performs that same computation under the left-to-right computation rule.

The technique had some drawbacks.
For completeness of the analysis phase, some very complex and technical conditions had to be satisfied.
The synthesis phase produced very complex logic programs.
It introduced new predicates that represent entire computation states (goals) observed in the symbolic evaluation tree and synthesized all transitions from each state to a next state.

Recently, in \cite{deschreyeanalysing2014} and \cite{nys2017abstract}, we revisited the CC technique.
Our motivation was not efficiency, but program analysis:
Coroutining computations --- the main type of computations dealt with by CC --- are notoriously difficult to analyze.
Many types of analyses of logic programs, such as termination analysis \cite{deschreyetermination1994}, have been developed for logic programs executed under the standard computation rule.
It is hard to adapt these techniques to a non-standard computation rule.
We aim to lift such analyses to programs with non-standard computation rules by revisiting CC.

In \cite{deschreyeanalysing2014}, our work focused on the analysis phase of CC.
We showed that the analysis could be reformulated and formalized using Abstract Conjunctive Partial Deduction (ACPD for short), a framework proposed by Leuschel in \cite{leuschelframework2004}.
In addition, we proposed a new abstraction, $\mathit{multi}$, to analyze computations with unboundedly growing goals.
This allowed us to analyze a diverse set of well-known coroutining programs and to compile these into programs executed under the standard computation rule.

In the current paper, we focus on the synthesis phase of CC.
Both in the original approach \cite{bruynooghe1989compiling} and in our revisited approach \cite{deschreyeanalysing2014}, the transformed program is a logic program, without delays or other non-declarative constructs.

However, the synthesis has some disadvantages.
First, the analysis phase analyzes the computational behavior of conjunctions of atoms.
Our instance of ACPD builds abstract derivation trees with conjunctions of (abstract) atoms in their roots and leaves.
To synthesize the branches of such trees into new clauses, renaming schemes are necessary to convert the conjunctions into atoms, so that they are then allowed to occur in the head of the Prolog clause.
If we compiled to a language with multi-headed rules, this complexity could be avoided.
CHR \cite{fruhwirth1998theory} is one such language.

Second, as mentioned above, the synthesis of a branch in the derivation tree compiles how some state in the computation is resolved to produce a new state.
As Prolog does not have a notion of ``state'', we need to encode entire states, both root states and those represented in the leaves, into new predicates.
This produces very complex predicates.
Their use could be avoided if we compiled to a language that keeps track of the current state for us,
and only requires us to express what changes with respect to the root state.
Again, CHR has a store that supports precisely this.

In this paper, we show how both these complexities can be overcome by synthesizing the resulting programs for the CC transformation in CHR, instead of Prolog.
The new synthesis has a number of advantages, including added insight into the computational process and more readable transformed code.
While we will assume the compilation target to be CHR(SWI-Prolog) for the sake of illustration,
this new technique also facilitates porting logic programs to CHR systems in other languages,
such as JCHR for Java or HaskellCHR for Haskell.
This work is ongoing and we do not yet have an automated system that synthesizes the CHR programs\footnote{An automated system which performs the analysis phase is available at \url{https://perswww.kuleuven.be/~u0055408/}}, but we aim to fully automate the transformation.

In what follows, we will assume the reader to be familiar with the basic concepts of partial deduction \cite{lloyd1991partial}, logic programming \cite{lloyd1987foundations} and CHR implementations \cite{schrijvers2008}.
Some familiarity with abstract interpretation \cite{bruynooghe1991practical} will be helpful, but it is not strictly required.

\section{An initial example: permutation sort}
\label{sec:permsort}
As in \cite{deschreyeanalysing2014}, we illustrate a use case of our technique by means of the permutation sort algorithm.
Listing \ref{lst:swipermsort} shows a naive version of this algorithm.
While we will retain the declarative perspective on sorting as creating an ordered permutation,
we would like to improve its \emph{implementation},
by using a computation rule that interleaves calls to $\mathit{perm}/2$ and $\mathit{ord}/1$,
so that ordering checks for elements of the permutation take place as soon as two list elements become ground.
Note that, in SWI-Prolog, $\mathit{select}/3$ is a built-in.

\begin{mylisting}
\noindent
\begin{minipage}[t]{0.5\linewidth}
\begin{Verbatim}[frame=none,numbers=left,framesep=3mm]
permsort(X,Y) :-
  perm(X,Y),
  ord(Y).

perm([],[]).
perm([X|Y],[U|V]) :-
  select(U,[X|Y],W),
  perm(W,V).
\end{Verbatim}
\end{minipage}\hfill
\begin{minipage}[t]{0.5\linewidth}
\begin{Verbatim}[frame=none,numbers=left,framesep=3mm]
ord([]).
ord([X]).
ord([X,Y|Z]) :-
  X =< Y,
  ord([Y|Z]).
\end{Verbatim}
\end{minipage}
\caption{Naive implementation of permutation sort in Prolog}
\label{lst:swipermsort}
\end{mylisting}

We briefly recall some key aspects of the CC transformation based on ACPD from \cite{deschreyeanalysing2014} and \cite{nys2017abstract}.
In the analysis phase, ACPD is applied with a specific abstract domain.
This abstract domain contains two collections of abstract variables $\{ a_i | i \in \mathbb{N}_0 \}$ and $\{ g_j | j \in \mathbb{N}_0 \}$.
The abstract variables $g_j$, $j \in \mathbb{N}_0$, represent ground terms in the concrete domain.
The variables $a_i$, $i \in \mathbb{N}_0$, represent any term of the concrete domain, including concrete variables.
If an abstract term, atom or conjunction of atoms contains some $a_i$ or $g_j$ several times (with the same index),
then the represented concrete terms, atoms or conjunctions of atoms all contain the same subterm at every position corresponding to the positions of the $a_i$ or $g_j$.
This represents aliasing between corresponding concrete terms.
Note that two variables $a_i$ and $g_i$, with the same subscript, are not considered to be aliased.

We also include concrete program constants in the abstract domain, such as the empty list, $[]$.
We treat these as $0$-arity functors.
This is different from \cite{deschreyeanalysing2014}, but it changes nothing about the procedure.

The desired computation rule is formalized by establishing a strict partial order on abstract atoms which occur during the analysis phase.
For permutation sort, the order consists of the pairs $\mathit{perm}(g_1,a_1) < \mathit{ord}(a_1)$, $\mathit{perm}(g_1,a_1) < \mathit{ord}([g_1|a_1])$ and $\mathit{ord}([g_1,g_2|a_1]) < \mathit{perm}(g_1,a_1)$.
We leave the details for the next section and restrict ourselves to saying that, if an abstract conjunction containing (a renamed variant of) a left-hand atom and (a renamed variant of) a right-hand atom from one of these pairs is encountered, the left-hand one will take precedence.
For instance, $\mathit{perm}(g_1,a_1)$ would take precedence over $\mathit{ord}(a_1)$, but $\mathit{ord}([g_1,g_2|a_1])$ would take precedence over $\mathit{perm}(g_1,a_1)$.
There are some additional details for computation rules, but we will postpone those until the next section.
Note that, if the desired computation rule cannot be expressed as a strict partial order, our current technique cannot be used.

The transformation requires a top level abstract goal, for which the analysis and transformation are performed.
For permutation sort, let $\mathit{permsort}(g_1, a_1)$ be that goal.

\begin{figure}
  \makebox[\textwidth][c]{
    \begin{minipage}[b]{.5\textwidth}
    \centering
    \includegraphics[scale=0.9]{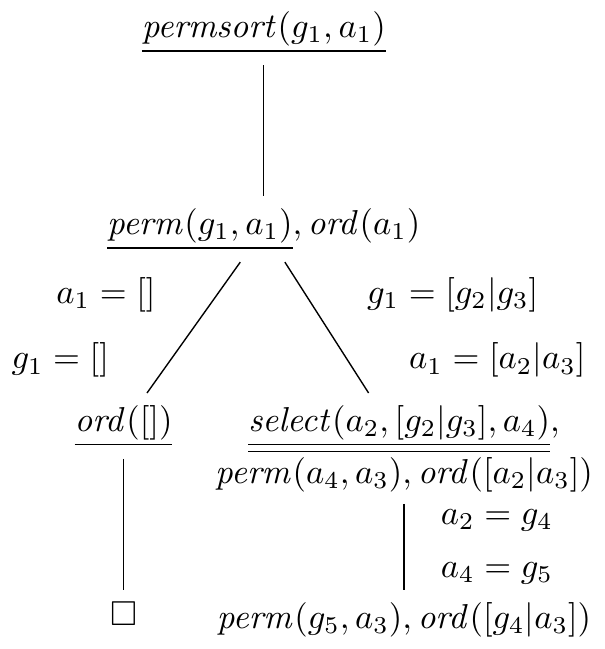}
    \caption{Abstract tree for $\mathit{permsort}(g_1,a_1)$}
    \label{abstractpermsort1}
    \end{minipage}
    \begin{minipage}[b]{.5\textwidth}
    \centering
    \includegraphics[scale=0.9]{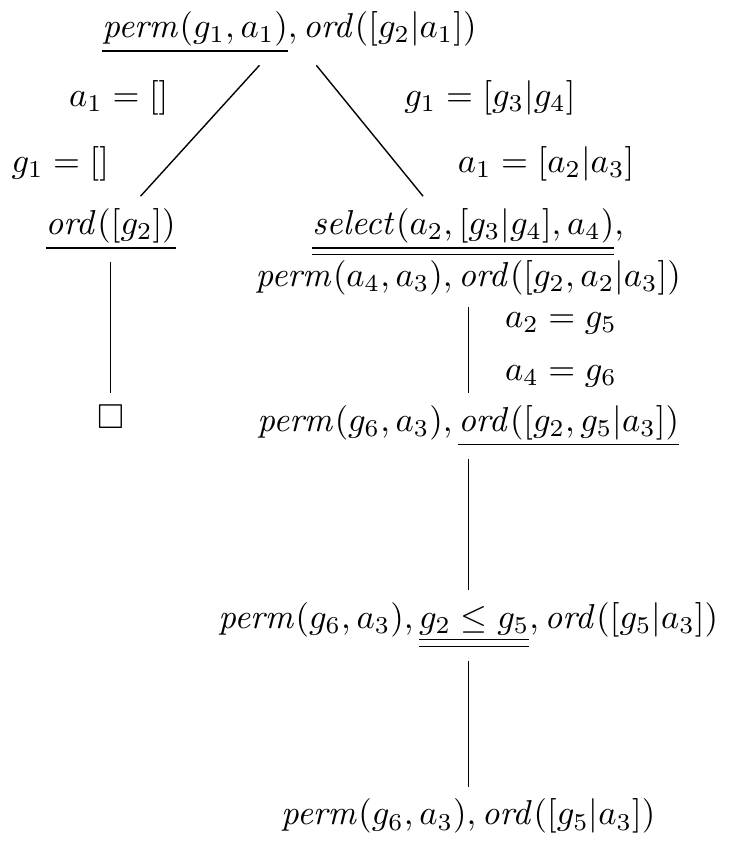}
    \caption{Abstract tree for $\mathit{perm}(g_1,a_1),\mathit{ord}([g_2|a_1])$}
    \label{abstractpermsort2}
    \end{minipage}
  }
\end{figure}

The analysis phase constructs a number of abstract derivation trees for a set $\mathcal{A}$ of abstract conjunctions.
In the example, $\mathit{permsort}(g_1,a_1)$ is the first of these conjunctions.
Starting from this goal, the analysis constructs the abstract trees in Figures \ref{abstractpermsort1} and \ref{abstractpermsort2}.
The set $\mathcal{A}$ consists of all the root nodes in these trees, so $\mathcal{A} = \{ permsort(g_1, a_1), \wedge(perm(g_1,a_1),ord([g_2|a_1])) \}$.
The set $\mathcal{A}$, with corresponding abstract trees, is said to be $\mathcal{A}$-closed, meaning that all the leaves in the trees are either success nodes, or are more specific than some element of $\mathcal{A}$.
Here, ``more specific'' is defined using an order on the abstract domain, consistent with set inclusion of the represented concrete terms, atoms and conjunctions.

\begin{figure}
  \makebox[\textwidth][c]{
    \begin{minipage}[b]{.5\textwidth}
    \centering
  \includegraphics[scale=0.9]{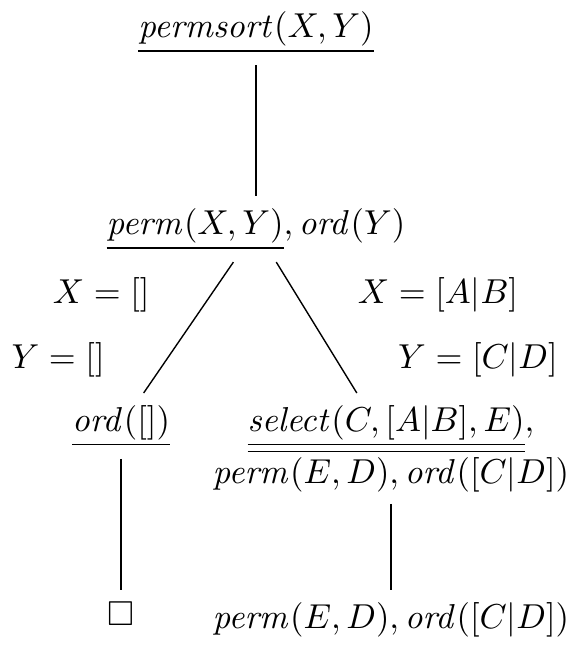}
  \caption{Concrete tree for $\mathit{permsort}(X,Y)$}
  \label{concretepermsort1}
    \end{minipage}
    \begin{minipage}[b]{.5\textwidth}
    \centering
  \includegraphics[scale=0.9]{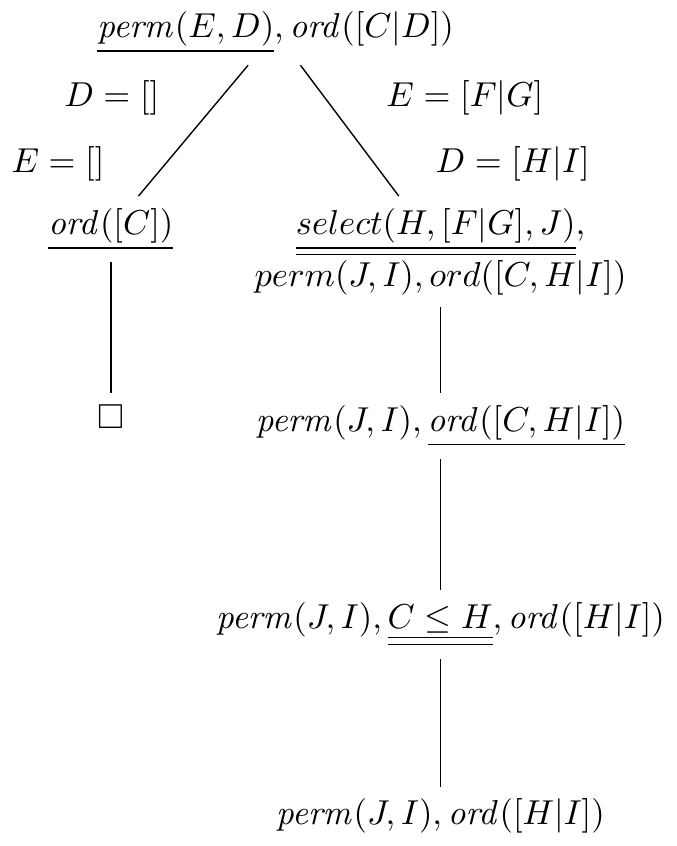}
  \caption{Concrete tree for $\mathit{perm}(E,D),\mathit{ord}([C|D])$}
  \label{concretepermsort2}
    \end{minipage}
  }
\end{figure}

In Figures \ref{abstractpermsort1} and \ref{abstractpermsort2}, underlined atoms are the atoms that are selected by the (non-standard) computation rule.
Atoms underlined twice are not only selected, but are also completely executed using abstract interpretation over the domain.
In the child node, these atoms have disappeared and any abstract bindings that the abstract interpretation of the abstract atom has produced are applied to the remaining atoms of the conjunction.
This feature is used for built-ins, but also for user-defined atoms that we do not want to transform (because their evaluation is not interleaved with that of the analyzed coroutines).
In the transformed program, we will simply rely on the original predicate definitions to solve such atoms.
Atoms underlined twice also bypass the strict partial order.
They could be included in the strict partial order, but to do so would complicate our representations of computation rules and would not provide us with any needed flexibility.
Instead, they are simply evaluated as soon as possible in left-to-right order.

After the analysis, the synthesis phase of \cite{deschreyeanalysing2014} starts by building a concrete derivation tree for every abstract tree in the analysis phase.
For the example, these trees are shown in Figures \ref{concretepermsort1} and \ref{concretepermsort2}.
The root of these trees is obtained from the root of the corresponding abstract tree by replacing $a_i$ and $g_j$ symbols by concrete variables.
The remainder of the trees is obtained by mimicking, over the concrete domain, the abstract resolution steps performed in the abstract tree, using the same clauses from the initial program.
In the concrete trees, atoms corresponding to atoms underlined twice in the abstract tree are not unfolded.
The atoms are removed from the goal and are considered residual.
This is also indicated by underlining these atoms twice.
A formal explanation of how concrete analysis trees are derived from abstract analysis trees can be found in \cite{nys2017abstract}.

Then, the actual (Prolog) code generation is performed.
For any non-failing branch of a concrete tree, a resultant is derived.
The head of the resultant is the conjunction in the root of the tree, with all bindings computed in the derivation applied to it.
The body is the leaf of the branch, preceded by all residual atoms on the branch.
In other words, it contains all atoms that correspond to double underlined abstract atoms, in addition to a conjunction that was not unfolded any further in this tree.
Note that the resultants are not necessarily Horn clauses: they can be multi-headed.
As an example, $\mathit{perm}([],[]) \wedge \mathit{ord}([C]) \leftarrow \mathit{true}$ is obtained from the leftmost branch of Figure \ref{concretepermsort2}.

Finally, conjunctions of atoms occurring in the heads of resultants are renamed to new predicates.
The same renaming is applied to their occurrences in the bodies of rules.
The resulting new clauses for permutation sort are shown in Listing \ref{lst:concretepermsort}.
Note how the meta-predicate $p1/2$ is used to group $\mathit{perm}/2$ and $\mathit{ord}/1$ atoms.
Also note that redundant structure could be removed from $\mathit{p1/2}$ as described in \cite{gallagher1990some}, but we have not done so to keep the mapping from the concrete trees to Listing \ref{lst:concretepermsort} as transparent as possible.

\begin{mylisting}
\noindent
\begin{minipage}[t]{0.5\linewidth}
\begin{Verbatim}[frame=none,numbers=left,framesep=3mm]
permsort([],[]).
permsort([A|B],[C|D]) :-
  select(C,[A|B],E),
  p1(perm(E,D),ord([C|D])).
\end{Verbatim}
\end{minipage}\hfill
\begin{minipage}[t]{0.5\linewidth}
\begin{Verbatim}[frame=none,numbers=left,framesep=3mm]
p1(perm([],[]),ord([C])).
p1(perm([F|G],[H|I]),ord([C,H|I])) :-
  select(H,[F|G],J),
  C =< H,
  p1(perm(J,I),ord([H|I])).
\end{Verbatim}
\end{minipage}
\caption{Prolog synthesis of permutation sort}
\label{lst:concretepermsort}
\end{mylisting}

In the current paper, we propose a new synthesis technique.
This new technique generates a CHR program which, under the refined operational semantics of CHR (used by implementations) produces equivalent answers to the original program.
It also evaluates constraints in the same order in which the abstract analysis evaluates corresponding abstract atoms.
An advantage of synthesizing to CHR is that conjunctions in $\mathcal{A}$ do not need to be renamed if they are independent.
We say that conjunctions $A$ and $B$ are independent if the intersection of the concrete instances of $A$ and the concrete instances of $B$ is empty.

For permutation sort, the synthesized CHR rules are in Listing \ref{lst:optimizedchrcoropermsort}.
The procedure for generating the CHR program is the following:
First, the non-built-in predicates of the initial program are partitioned into two sets: those that will be CHR constraints in the synthesized program and those that will be built-in constraints (in the terminology of CHR, this means that they will remain Prolog predicates which may or may not be user-defined).
The partitioning criterion is easy: Predicates that are fully abstractly evaluated in the abstract trees are built-in constraints, those that are abstractly unfolded in the trees become CHR constraints.
We require the abstract analysis to be consistent in the choice between these two options.
For permutation sort, all user-defined predicates become CHR constraints.

Next, for every non-failing branch in a (concrete) derivation tree, we synthesize one simplification rule.
The head of this rule consists of all atoms in the root of the tree that are unfolded in the branch, after applying a subset of the unifications performed along the branch.
The body of the rule consists of residual atoms evaluated along the branch, the remaining unifications performed along the branch and all the atoms in the leaf of the branch which are newly introduced with respect to the root.
Precisely which unifications are applied to the head and in the body is determined as follows:
For every binding in the derivation that binds a variable in an unfolded atom of the root, we check the corresponding abstract tree to see whether applying the concrete binding could further instantiate the concrete term.
If so, we add the binding as an explicit unification in the body of the rule.
If not, we apply the binding to the head of the CHR rule.
The reason for this distinction is that CHR programs, unlike Prolog programs, use \emph{one-way} unification.
In other words, a rule can only fire if its head is at least as general as the set of constraints which causes the rule to fire\footnote{We could therefore also move all unifications to the rule body, synthesizing CHR rules in head normal form, but we prefer our rules to be as concise as possible.}.
In permutation sort, this gives the rules in Listing \ref{lst:optimizedchrcoropermsort}.

Finally, we order the generated CHR rules, using the strict partial order on abstract atoms mentioned earlier.
If a CHR rule contains a head constraint, such that the corresponding abstract atom in the abstract tree it was generated from takes precedence over the abstract atoms (concretized as constraints) in the head of another rule, then the former rule must precede the latter.
If both rule heads concretize the same top-priority abstract atom, the abstract atom with the second highest priority is considered, etc.
This ensures that the priority of unfolding atoms, expressed in the computation rule, is maintained in the CHR program.
For permutation sort, all orderings of the rules lead to an equivalent program.

\begin{mylisting}
\noindent
\begin{minipage}[t]{0.5\linewidth}
\begin{Verbatim}[frame=none,numbers=left,framesep=3mm]
permsort([],X) <=> X = [].
permsort([A|B],Y) <=>
  Y = [C|D],
  select(C,[A|B],E),
  perm(E,D),
  ord([C|D]).
\end{Verbatim}
\end{minipage}\hfill
\begin{minipage}[t]{0.5\linewidth}
\begin{Verbatim}[frame=none,numbers=left,framesep=3mm]
perm([],D), ord([C|D]) <=> D = [].
perm([F|G],D), ord([C|D]) <=>
  D = [H|I],
  select(H,[F|G],J),
  C =< H,
  perm(J,I),
  ord([H|I]).
\end{Verbatim}
\end{minipage}
\caption{CHR permutation sort}
\label{lst:optimizedchrcoropermsort}
\end{mylisting}

Comparing Listing \ref{lst:concretepermsort} with Listing \ref{lst:optimizedchrcoropermsort},
permutation sort already allows us to draw some conclusions in comparing the two synthesis approaches.
First, there is no more need to rename conjunctions of atoms to a new predicate, in order to obtain the clause format.

The second improvement is that, where the Prolog synthesis needs to represent transitions of a complete goal (conjunction) in the root of a tree to the complete goal in the leaf of the tree, the CHR synthesis only needs to focus on the unfolded atoms of the root.
For permutation sort, however, the Prolog and CHR syntheses are very similar, because all constraints are rewritten simultaneously.
Our second example will highlight both advantages.

\section{Incorporating the $\mathit{multi}$ abstraction: confused queens}
\label{sec:confused}
Permutation sort is a very easy example for CC.
For most other coroutining programs, compiling them with the CC transformation involves an extra abstraction.
We call this the $\mathit{multi}$ abstraction \cite{deschreyeanalysing2014}.
In this section, we study the transformation again, and in particular the synthesis to CHR, but for a more representative example, requiring the $\mathit{multi}$ abstraction: the confused queens problem.

We will first formalize the computation rule,
so that we can precisely express the unfoldings performed in our analysis of confused queens without showing every abstract analysis tree.
We will then apply the transformation.
We will see that it needs some refinements, which will be introduced in Section \ref{sec:refinement}.

We begin by briefly introducing some core concepts.
If a concept related to the abstract domain $\mathit{ADom_P}$ is not specified, the reader should assume that it is analogous to its counterpart in the concrete domain $\mathit{Dom_P}$.
For a complete formalization, see \cite{nys2017abstract}.
The abstract domain contains the two aforementioned sets of abstract variables, whose union is denoted $\mathit{AVar_P}$.
Based on these, there is a corresponding set of abstract terms, $\mathit{ATerm_P}$,
which consists of the terms constructed from $\mathit{AVar_P}$ and function symbols and constant symbols occurring in the concrete program $P$.
$\mathit{AAtom_P}$ denotes the set of abstract atoms, i.e. the atoms which can be constructed from $\mathit{ATerm_P}$ and predicate symbols occurring in $P$.
$\mathit{AConAtom_P}$ denotes the set of conjunctions of elements of $\mathit{AAtom_P}$.
Let $\mathit{AAtom_{P/\approx}}$ and $\mathit{AConAtom_{P/\approx}}$ denote equivalence classes of abstract atoms and conjunctions, respectively.
Two abstract atoms (or conjunctions) $A$ and $B$ are equivalent, denoted $A \approx B$,
if and only if there are abstract substitutions $\theta_1$ and $\theta_2$ such that $A \theta_1 = B$ and $B \theta_2 = A$.
For instance, $p1(g_1,a_1) \approx p1(g_2,a_2)$, but $f(g_1,a_1) \not\approx f(g_2,g_3)$.
In what follows, we will often use an abstract atom as a representative of its equivalence class.
We assume the meaning of any overloaded concepts related to abstract atoms and equivalence classes to be clear from the context.

Finally, let $\gamma: \mathit{ADom_P} \rightarrow 2^{\mathit{Dom_P}}$  be the \importantterm{concretization function}, which maps elements of the abstract domain to their concrete denotation.

\begin{definition}[Instantiation-based computation rule]
\label{def:computation-rule}
An instantiation-based computation rule for a program $P$ is a strict partial order $<$ (``is less than'') on $\mathit{AAtom_{P/\approx}}$, such that $\gamma(s_1) \subset \gamma(s_2)$ implies $s_2 \nless s_1$, where $\subset$ denotes strict set inclusion.
\end{definition}

An instantiation-based computation rule expresses which atom is selected from an abstract conjunction, either for expansion or full evaluation.
Our technique requires that an instantiation-based computation rule can specify the desired control flow.

\begin{definition}[Selection by an instantiation-based computation rule]
\label{def:selection-by-computation-rule}
Let $A \in \mathit{ACon_{P/\approx}}$.
Then, the abstract atom selected from $A$ by $<$ is the leftmost atom $b$, such that $\forall c \in A: c \not\approx b \Rightarrow b < c$.
If there is no such atom $b$, no atom is selected from $A$.
\end{definition}

To formalize certain properties of a computation rule, we require the notion of a reachable state.
Strictly speaking, a reachable state is an equivalence class of abstract conjunctions.
We will define it as a representative $A$.
The reachable state proper is then its equivalence class.

\begin{definition}[Representative of a reachable state]
An abstract conjunction $A$ is a representative of a reachable state under the computation rule $<$ for the program $P$ with initial query $q$ if either:
\begin{itemize}
  \item $A = q$;
  \item $A$ is the abstract resolvent of $b$ in $B$ with $\mathit{AC}$, where $B$ is a reachable state, $\mathit{AC}$ is an abstraction of a Horn clause in $P$ and $b$ is the atom selected by $<$ from $B$;
  \item $A = c_1,\ldots,c_{i-1},c_{i+1},\ldots,c_m \theta$, where:
  \begin{itemize}
    \item a representative of a reachable state $B = c_1,\ldots,c_{i-1},c_i,c_{i+1},\ldots,c_m$ exists;
    \item $c_i$ is the atom selected by $<$ from $B$;
    \item $c_i$ is fully abstractly evaluated in $P$, yielding the abstract unifier $\theta$.
  \end{itemize}
\end{itemize}
\end{definition}

A computation rule is \importantterm{complete} if it selects an abstract atom in every reachable state.
We require that the computation rule to be analyzed is complete.
We do not, however, expect users to supply a formal, complete computation rule before beginning the program analysis.
Instead, we use Algorithm \ref{alg} to construct a set of partial ordering pairs, $\mathit{CR}$, during program analysis:

\begin{algorithm}
\caption{Constructing a complete computation rule}
\label{alg}
\begin{algorithmic}
\State $\mathit{CR} \gets \emptyset$
\Loop
  \State let $c_1,\ldots,c_n$ be the next goal in the analysis tree
  \If{$\mathit{CR}$, Definition \ref{def:computation-rule} and Definition \ref{def:selection-by-computation-rule} restrict the choice to a single $c_i, i \in \{1,\ldots,n\}$}
    \State proceed without asking for user input
  \Else
    \State let the user select an atom $c_j$ from $c_1,\ldots,c_n$
    \If{the user's selection is consistent with $\mathit{CR}$, Definition \ref{def:computation-rule} and Definition \ref{def:selection-by-computation-rule}}
      \State $\mathit{CR} \gets \mathit{CR} \cup \{c_j < c_k | k \in \{ 1,\ldots,n \}, c_j$ is not equivalent to $c_k \}$
    \Else
      \State reject the selection of $c_j$
    \EndIf
  \EndIf
\EndLoop
\end{algorithmic}
\end{algorithm}

For the sake of brevity, we will not list the entire rule.
Instead, we will list a generating set.
Finally, we assume that fully evaluated atoms are dealt with in left-to-right order, before unfolded atoms.

We are now ready to introduce the confused queens problem and the computation rule that we use to analyze it.
The confused queens problem is a variation on the well-known $N$-queens problem.
Like in $N$-queens, the goal is to arrange $N$ queens on a chess board of size $N \times N$.
Unlike in the $N$-queens problem, for confused queens, every pair of queens should be on the same row or on the same diagonal.
Each piece is still placed in a different column, which is implicit in the use of a list data structure.
Prolog code for the program (using a standard computation rule) is shown in Listing \ref{lst:swiconfused}.

\begin{mylisting}
\noindent
\begin{minipage}[t]{0.5\linewidth}
\begin{Verbatim}[frame=none,numbers=left,framesep=3mm]
cqueens(N,D) :-
  genlist(N,L),
  draw(N,L,D),
  confused(D).

genlist(N,L) :-
  N >= 1,
  genlist_acc(N,[],L).
genlist_acc(N,Acc,L) :-
  N > 1,
  Nmin is N-1,
  genlist_acc(Nmin,[N|Acc],L).
genlist_acc(1,Acc,[1|Acc]).

draw(0,_,[]).
draw(N,L,[E|R]) :-
  N > 0,
  Nmin is N - 1,
  member(E,L),
  draw(Nmin,L,R).
\end{Verbatim}
\end{minipage}\hfill
\begin{minipage}[t]{0.5\linewidth}
\begin{Verbatim}[frame=none,numbers=left,framesep=3mm]
confused([]).
confused([_X]).
confused([A,B|C]) :-
  attack_all(A,1,[B|C]),
  confused([B|C]).

attack_all(_,_,[]).
attack_all(A,Off,[B|C]) :-
  Offplus is Off + 1,
  attack(A,Off,B),
  attack_all(A,Offplus,C).

attack(A,_,A).
attack(A,Off,B) :-
  Diff is A - B,
  abs(Diff,Off).
\end{Verbatim}
\end{minipage}
\caption{Prolog implementation of confused queens}
\label{lst:swiconfused}
\end{mylisting}

The aspect that makes permutation sort easy with respect to CC is that the number of abstract atoms in any node of any abstract derivation tree is bounded (at most three).
For confused queens, and most other examples, there is no such bound.
Of course, in any concrete evaluation of $\mathit{cqueens}(N,D)$, with $N$ a natural number, the number of atoms that can appear in a goal is bounded.
But that bound increases with $N$, so that for an ACPD analysis in the style of Section \ref{sec:permsort}, which is a safe approximation of every concrete computation, there cannot be a bound. To deal with this, \cite{deschreyeanalysing2014} introduces the $\mathit{multi}$ abstraction. 

We analyze the program for the abstract goal $\mathit{cqueens}(g_1, a_1)$.
The generating set for the rule is:\\[1em]
$\{ (\mathit{attack\_all}(g_1,g_2,[g_3|a_1]),\mathit{attack\_all}(g_1,g_2,a_1)),
    (\mathit{attack\_all}(g_1,g_2,[g_3|a_1]),\mathit{confused}([g_1|a_1])),\\
    (\mathit{attack\_all}(g_1,g_2,[g_3|a_1]),\mathit{draw}(g_1,g_2,a_1)),
    (\mathit{confused}([g_1]),\mathit{attack\_all}(g_1,g_2,[])),\\
    (\mathit{confused}([g_1,g_2|a_1]),\mathit{attack\_all}(g_1,g_2,[g_3|a_1])),
    (\mathit{confused}([g_1,g_2|a_1]),\mathit{draw}(g_1,g_2,a_1)),\\
    (\mathit{draw}(g_1,g_2,a_1),\mathit{confused}(a_1)),
    (\mathit{draw}(g_1,g_2,a_1),\mathit{confused}([g_1|a_1])),\\
    (\mathit{draw}(g_1,g_2,a_1),\mathit{attack\_all}(g_1,g_2,a_1)) \}$\\

Consider the following abstract goals that occur in the abstract trees built for this top level goal using the given computation rule:

\begin{itemize}
\item $\mathit{draw}(g_1,g_2,a_1), \mathit{attack\_all}(g_4,g_5,a_1), \allowbreak \mathit{attack\_all}(g_6, g_7, [g_3|a_1]), \allowbreak \mathit{confused}([g_3|a_1])$
\item $\mathit{draw}(g_1,g_2,a_1), \mathit{attack\_all}(g_4,g_5,a_1), \allowbreak \mathit{attack\_all}(g_8,g_9,a_1), \allowbreak \mathit{attack\_all}(g_6,g_7,[g_3|a_1]), \break \mathit{confused}([g_3|a_1])$

\item $\mathit{draw}(g_1,g_2,a_1), \mathit{attack\_all}(g_4,g_5,a_1),\allowbreak \mathit{attack\_all}(g_8,g_{11},a_1),\allowbreak \mathit{attack\_all}(g_6,g_7,[g_3|a_1]),\break \mathit{attack\_all}(g_{10},g_{11},[g_3|a_1]),\allowbreak \mathit{confused}([g_3|a_1])$
\end{itemize}

These three abstract goals are very similar.
They have identical $\mathit{draw}/2$ and $\mathit{confused}/1$ abstract atoms.
They all have two types of $\mathit{attack\_all}/3$ abstract atoms: one type with an $a_1$ as the third argument and another with $[g_4|a_1]$ as the third argument.
The difference between the three abstract goals is the number of abstract atoms of each type which each goal contains: $1$ and $1$ for the first goal, $2$ and $1$ for the second, $2$ and $2$ for the third.

In \cite{deschreyeanalysing2014}, we introduced a new abstraction, $\mathit{multi}$.
This new abstraction makes it possible to generalize the three goals above to the following new abstract goal:
$\mathit{draw}(g_1,g_2,a_1), \allowbreak \mathit{multi}(\mathit{attack\_all}(g_i, g_j, a_1)), \allowbreak  \mathit{multi}(\mathit{attack\_all}(\allowbreak g_k,g_l,[g_4|a_1])), \allowbreak \mathit{confused}([g_4|a_1])$\footnote{Note that we use a more lightweight notation here than in \cite{deschreyeanalysing2014} for the sake of presentation. For confused queens, the lightweight notation is sufficient.}.

The semantics of $\mathit{multi}(\mathit{attack\_all}(g_i, g_j, a_1))$ is as follows:
the abstraction represents all conjunctions of one or more $\mathit{attack\_all}(g_i, g_j, a_1)$ abstract atoms,
where the first arguments of consecutive conjuncts are not aliased,
where the same holds for the second arguments,
but in each represented conjunct, the last argument is identical.
The $\mathit{multi}$ abstraction is included in $\mathit{ADom_P}$.

We refer to \cite{nys2017abstract} for a more formal account of $\mathit{multi}$ and the operations which can be applied to it: generalization and unfolding.
A simplified and informal description of these operations will do for the current discussion, as the synthesis --- not the analysis --- is the focus of the current work.
Generalization can take two forms.
The first form replaces an abstract conjunction, or several syntactically consecutive, equivalent abstract conjunctions.
Their replacement is a $\mathit{multi}$ abstraction whose argument has the form of a renamed variant of the abstracted conjunction(s), but which may use symbolic subscripts to indicate that abstract variables occurring in consecutive conjunctions are not aliased.
In this way, the conjunction containing the $\mathit{multi}$ abstraction represents all abstract goals with one or more of such conjunctions in the syntactic position of the $\mathit{multi}$ abstraction.
The second form removes a single renamed variant of the argument of an existing, syntactically adjacent $\mathit{multi}$ abstraction.
This should be understood as a generalization from ``two or more'' equivalent abstract conjunctions to ``one or more'' equivalent conjunctions.
For example, $\mathit{atom}(g_1,a_1),\mathit{multi}(\mathit{atom}(g_i,a_j))$ can be generalized to $\mathit{multi}(\mathit{atom}(g_i,a_j))$.

In this work, we simply assume that generalization is applied in such a way that $\mathcal{A}$ can be closed.
Until recently, this meant performing part of the analysis manually and grouping equivalent subconjunctions.
In an upcoming release of the implementation, we exploit the fact that these subconjunctions are introduced by evaluating certain abstract atoms recursively, but not tail-recursively, under the computation rule.
Subconjunctions which are grouped are then unfinished computations indirectly introduced by a single, top-level, recursive abstract atom.
We leave the details of this analysis for future work.

Unfold of $\mathit{multi}$ describes what it means for an atom in a $\mathit{multi}$ to be selected and unfolded.
It makes a case split.
Either the $\mathit{multi}$ represents one abstract conjunction, or it represents more than one abstract conjunction.
In the first case, the $\mathit{multi}$ disappears from the goal and is replaced by a conjunction equivalent to the argument of the $\mathit{multi}$ (with symbolic indices replaced by fresh indices).
In the second, the $\mathit{multi}$ remains in the abstract goal, but one extra conjunction, again equivalent to the argument of the $\mathit{multi}$ but without symbolic indices, is added to the goal.
In both cases, atoms from the newly introduced abstract conjunction can then be unfolded.

\begin{figure}
  \makebox[\textwidth]{
    \begin{minipage}[b]{.5\textwidth}
    \centering
    \includegraphics[scale=0.9]{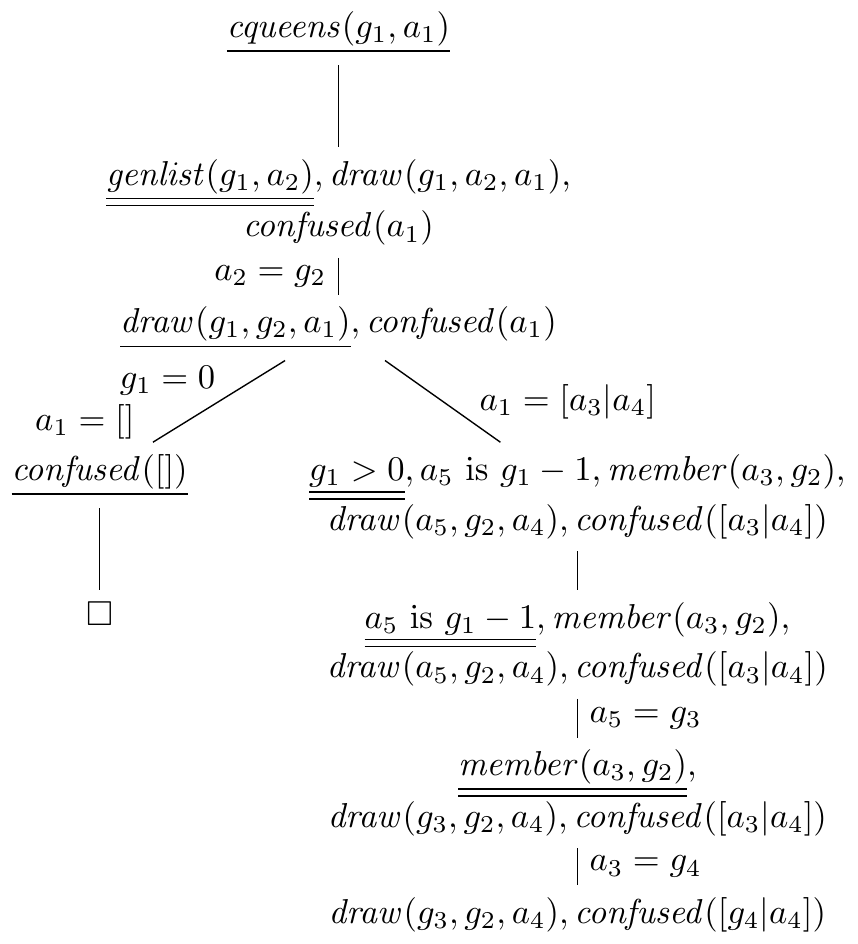}
    \caption{First abstract tree for $\mathit{cqueens}(g_1,a_1)$.}
    \label{fig:abstracttreecqueens1}
    \end{minipage}
    \begin{minipage}[b]{.5\textwidth}
    \hspace*{-2.2cm}
    \includegraphics[scale=0.9]{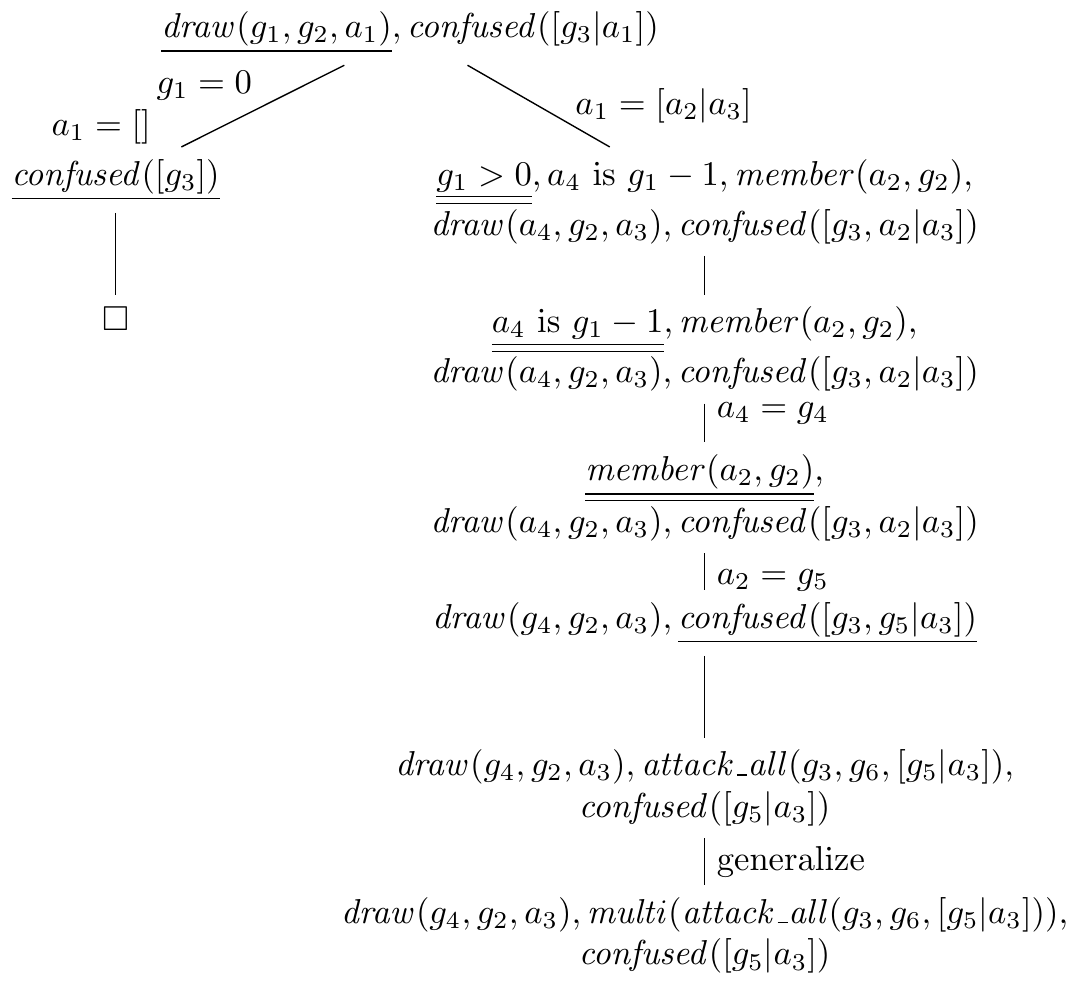}
    \caption{Second abstract tree.}
    \label{fig:abstracttreecqueens2}
    \end{minipage}
  }
\end{figure}

\begin{figure}
  \centering
\makebox[\textwidth][c]{
  \includegraphics[scale=0.9]{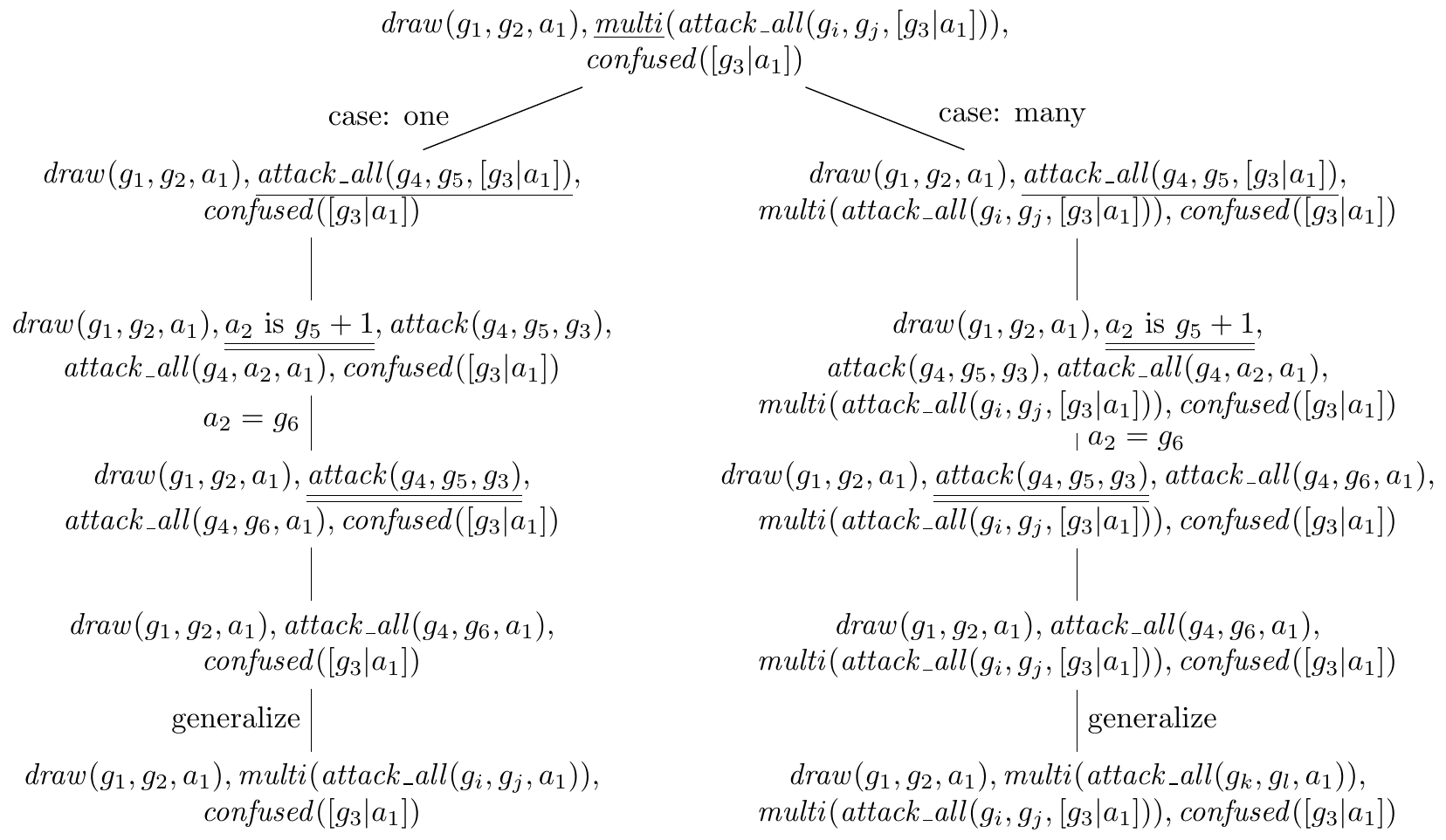}
}
  \caption{Third abstract tree.}
  \label{fig:abstracttreecqueens3}
\end{figure}

The trees in Figures \ref{fig:abstracttreecqueens1}, \ref{fig:abstracttreecqueens2} and \ref{fig:abstracttreecqueens3} show the first, second and third abstract derivation trees built for the ACPD analysis for $\mathit{cqueens}(g_1,a_1)$.
The remaining abstract trees, for an $\mathcal{A}$-closed ACPD analysis for $\mathit{cqueens}(g_1,a_1)$ are provided in appendix.
They can also be reproduced using the strict partial order given above.
The concrete counterpart to Figure \ref{fig:abstracttreecqueens3} is shown in Figure \ref{fig:concretetreecqueens3a}.
It illustrates how the $\mathit{multi}$ abstraction is represented at the concrete level.
The remaining concrete trees are also in appendix.
After generating the concrete trees, resultants for the branches can be computed.
The resultants for Figures \ref{fig:abstracttreecqueens1} through \ref{fig:abstracttreecqueens3} are as follows (with $\mathit{attack\_all}$ renamed to $\mathit{atk\_all}$ for layout purposes):
\begin{itemize}
\item $\mathit{cqueens}(0,[]).$
\item $\mathit{cqueens}(A,[D|E]) \leftarrow\\ \mathit{genlist}(A,C), \allowbreak A > 0, \allowbreak F\text{ is }A - 1, \allowbreak \mathit{member}(D,C), \allowbreak \mathit{draw}(F,C,E), \allowbreak \mathit{confused}([D|E]).$
\item $\mathit{draw}(0,X,[]), \mathit{confused}([Y]).$
\item $\mathit{draw}(A,B,[E|F]), \mathit{confused}([D,E|F]) \leftarrow\\ A \geq 0, G \text{ is } A-1, \mathit{member}(E,B), \mathit{draw}(G,B,F), \mathit{multi}([\mathit{atk\_all}(D,1,[E|F])]), \mathit{confused}([E|F]).$
\item $\mathit{draw}(A,B,C), \mathit{multi}([\mathit{atk\_all}(D,E,[F|C])]), \mathit{confused}([F|C]) \leftarrow\\ H \text{ is } E + 1, \mathit{attack}(D,E,F), \mathit{draw}(A,B,C), \mathit{multi}([\mathit{atk\_all}(D,H,C)|I]), \mathit{confused}([F|C])$
\item $\mathit{draw}(A,B,C), \mathit{multi}([\mathit{atk\_all}(D,E,[F|C]),\mathit{atk\_all}(H,I,[F|C])|J)]), \mathit{confused}([F|C]) \leftarrow\\ K \text{ is } E + 1, \mathit{attack}(D,E,F), \mathit{draw}(A,B,C), \mathit{multi}([\mathit{atk\_all}(D,K,C)|L]),\\\mathit{multi}([\mathit{atk\_all}(H,I,[F|C])|J]), \mathit{confused}([F|C])$
\end{itemize}

\begin{figure}[h!]
  \centering
\makebox[\textwidth][c]{
  \includegraphics[scale=0.9]{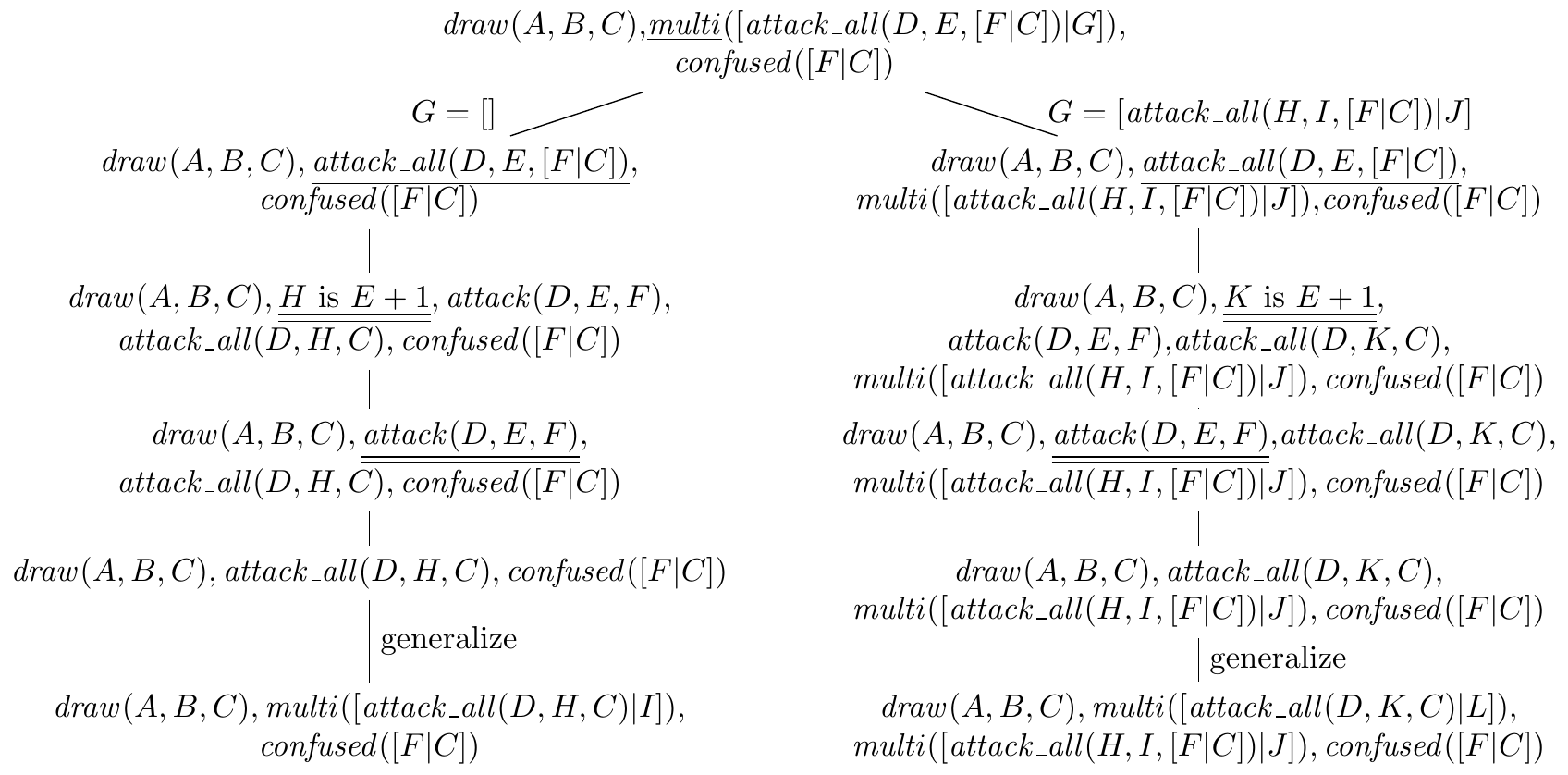}
}
  \caption{Concrete tree corresponding to Figure \ref{fig:abstracttreecqueens3}.}
  \label{fig:concretetreecqueens3a}
\end{figure}

Finally, when all resultants of all trees have been computed, renaming transformations are applied to obtain Prolog clauses.
This also ensures independence of the concrete goals.
The resulting clauses corresponding to the resultants for the first three trees are shown in Listing \ref{lst:firstresultantsconcreteconfusedqueens}.
The full synthesized Prolog program is in appendix.

\begin{mylisting}
\noindent
\begin{minipage}[t]{0.5\linewidth}
\begin{Verbatim}[frame=none,numbers=left,framesep=3mm]
cqueens(0,[]).
cqueens(A,[D|E]) :-
   genlist(A,C),
   A > 0,
   F is A - 1,
   member(D,C),
   a(draw(F,C,E),
     confused([D|E])).
a(draw(0,_,[]),confused([_])).
a(draw(A,B,[E|F]),
  confused([D,E|F])) :-
  A >= 0,
  G is A - 1,
  member(E,B),
  b(draw(G,B,F),
    multi([attack_all(D,1,[E|F])]),
    confused([E|F])).
\end{Verbatim}
\end{minipage}\hfill
\begin{minipage}[t]{0.5\linewidth}
\begin{Verbatim}[frame=none,numbers=left,framesep=3mm]
b(draw(A,B,C),
  multi([attack_all(D,E,[F|C])]),
  confused([F|C])) :-
  H is E + 1,
  attack(D,E,F),
  c(draw(A,B,C),
    multi([attack_all(D,H,C)]),
    confused([F|C])).
b(draw(A,B,C),
  multi([attack_all(D,E,[F|C]),
  attack_all(H,I,[F|C])|J]),
  confused([F|C])) :-
  K is E + 1,
  attack(D,E,F),
  d(draw(A,B,C),
    multi([attack_all(D,K,C)]),
    multi([attack_all(H,I,[F|C])|J]),
    confused([F|C])).
\end{Verbatim}
\end{minipage}
\caption{Renamed clauses for the first six resultants of confused queens}
\label{lst:firstresultantsconcreteconfusedqueens}
\end{mylisting}

Note that these clauses become rather complex.
They explicitly describe every possible transition from one concrete goal corresponding to an element of $\mathcal{A}$ to another.
This obscures the actual logic behind the transition, i.e. the \emph{change} that occurs between two states.
We are now in a position to show that the synthesis to CHR, as described in Section \ref{sec:permsort}, expresses this logic more clearly.

From the first branch in Figure \ref{fig:abstracttreecqueens1}, we generate:\\
\begin{minipage}{\textwidth}
\vspace{.2cm}
\begin{minipage}[t]{0.80\textwidth}\begin{flushleft}
\begin{verbatim}
cqueens(0,B) <=> genlist(0,C), B = [], lock.
\end{verbatim}
\end{flushleft}\end{minipage}\begin{minipage}[t]{0.2\textwidth}
\begin{flushright}(1)
\end{flushright}
\end{minipage}
\end{minipage}

\vspace{.2cm}
From the second branch:\\
\begin{minipage}{\textwidth}
\vspace{.2cm}
\begin{minipage}[t]{0.80\textwidth}\begin{flushleft}
\begin{verbatim}
cqueens(A, B) <=> B = [E|F], genlist(A,C), A > 0, D is A-1, member(E,C),
  draw(D,C,F), confused([E|F]), lock.
\end{verbatim}
\end{flushleft}\end{minipage}\begin{minipage}[t]{0.2\textwidth}
\begin{flushright}(2)
\end{flushright}
\end{minipage}
\vspace{.2cm}
\end{minipage}

The purpose of the lock constraint is to prevent a transition from firing at certain points during program execution.
The refined operational semantics of CHR dictates that atoms are evaluated and constraints are added to the store from left to right.
This means that a rule could fire \emph{while} another rule is being applied, raising an issue for our transformation:
rules represent transitions between instances of abstract states in $\mathcal{A}$.
Thus, they should only fire when the contents of the store are an instance of an abstract state in $\mathcal{A}$.
The $\mathit{lock}/0$ ensures that new rules only fire when an entire new state is reached, because it is always required for a state transition and because it is always added last.

The left branch of the tree in Figure \ref{fig:abstracttreecqueens2} yields:\\
\begin{minipage}{\textwidth}
\vspace{.2cm}
\begin{minipage}[t]{0.80\textwidth}\begin{flushleft}
\begin{verbatim}
draw(0,B,C), confused([D|C]), lock <=> C = [], lock.
\end{verbatim}
\end{flushleft}\end{minipage}\begin{minipage}[t]{0.2\textwidth}
\begin{flushright}(3)
\end{flushright}
\end{minipage}
\end{minipage}

\vspace{.2cm}
The right branch yields:\\
\begin{minipage}{\textwidth}
\vspace{.2cm}
\begin{minipage}[t]{0.80\textwidth}\begin{flushleft}
\begin{verbatim}
draw(A,B,C), confused([D|C]), lock <=> C = [E|F], A>0, G is A-1,
  member(E,B), draw(G,B,F), attack_all(D,1,[E|F]), confused([E|F]), lock.
\end{verbatim}
\end{flushleft}\end{minipage}\begin{minipage}[t]{0.2\textwidth}
\begin{flushright}(4)
\end{flushright}
\end{minipage}
\vspace{.2cm}
\end{minipage}

Note that, unlike in the Prolog synthesis, we do not introduce a representation of $\mathit{multi}$ in the body of the CHR rule.
We rely on the CHR store to accumulate multiple $\mathit{attack\_all}/3$ constraints.

From the left branch of the tree in Figure \ref{fig:abstracttreecqueens3}, we generate the following rule:\\
\begin{minipage}{\textwidth}
\vspace{.2cm}
\begin{minipage}[t]{0.80\textwidth}\begin{flushleft}
\begin{verbatim}
attack_all(D,E,[F|C]), lock <=>
  H is E + 1, attack(D,E,F), attack_all(D,H,C), lock.
\end{verbatim}
\end{flushleft}\end{minipage}\begin{minipage}[t]{0.2\textwidth}
\begin{flushright}(5)
\end{flushright}
\end{minipage}
\vspace{.2cm}
\end{minipage}

Again, we do not synthesize the unfolding of $\mathit{multi}$.
We only consider the unfolding of an $\mathit{attack\_all}/3$ constraint which is present in the store.

For the right branch of the tree in Figure \ref{fig:abstracttreecqueens3}, the generated rule is identical.
There are two leaf nodes in the abstract tree, but this is only due to the case split on $\mathit{multi}$.
The unfoldings and full evaluations of abstract atoms required to go from the root to a leaf node are identical for both branches.
In the CHR synthesis, there is no case split: a specific number of concrete instances of $\mathit{attack\_all}(g_i,g_j,[g_4|a_1])$ is in the constraint store.
That is, the two branches of Figure \ref{fig:concretetreecqueens3a} are mutually exclusive.
Therefore, it is not necessary to add an additional rule to the synthesized program.
In the Prolog synthesis, we do get a different synthesized clause for each of these branches, because the context of the derivation is different in each branch and the Prolog synthesis encodes the context in its new predicates.
Because rules do not need to be repeated and because they only express the change between states, the CHR synthesis is significantly more parsimonious than the Prolog synthesis.
In fact, several more branches in the full program analysis generate rule (5).

As a result, only the following new rule is synthesized from the remaining trees:\\
\begin{minipage}{\textwidth}
\vspace{.2cm}
\begin{minipage}[t]{0.80\textwidth}\begin{flushleft}
\begin{verbatim}
attack_all(X,Y,[]), lock <=> lock.
\end{verbatim}
\end{flushleft}\end{minipage}\begin{minipage}[t]{0.2\textwidth}
\begin{flushright}(6)
\end{flushright}
\end{minipage}
\end{minipage}

\vspace{.2cm}
Finally, a housekeeping rule is required:\\
\begin{minipage}{\textwidth}
\vspace{.2cm}
\begin{minipage}[t]{0.80\textwidth}\begin{flushleft}
\begin{verbatim}
lock <=> true.
\end{verbatim}
\end{flushleft}\end{minipage}\begin{minipage}[t]{0.2\textwidth}
\begin{flushright}(7)
\end{flushright}
\end{minipage}
\end{minipage}

\vspace{.2cm}
We always add rule (7) at the end of a synthesized program so that the $\mathit{lock}/0$ constraint is discarded in an otherwise empty state.

After synthesizing all the rules, we order them based on the strict partial ordering on abstract atoms as before.
The resulting program works as intended for the expected queries.
It is shown in Listing \ref{lst:unsafecqueens}.
The complete Prolog program, which is provided in appendix, contains nearly twice as many lines, and the lines themselves are significantly longer.

\begin{mylisting}[p]
\noindent
\begin{Verbatim}
:- use_module(library(chr)).
:- chr_constraint cqueens/2, draw/3, confused/1, attack_all/3, lock/0.
genlist(N,L) :- N >= 1, genlist_acc(N,[],L).
genlist_acc(N,Acc,L) :- N > 1, Nmin is N-1, genlist_acc(Nmin,[N|Acc],L).
genlist_acc(N,Acc,[1|Acc]) :- N is 1. 

attack(A,_,A).
attack(A,Offset,B) :- Diff is A - B, abs(Diff,Offset).

cqueens(0, B) <=> genlist(0,C), B = [].
cqueens(A, B) <=>
  B = [E|F],
  genlist(A,C),
  A > 0,
  D is A-1,
  member(E,C),
  draw(D,C,F),
  confused([E|F]),
  lock.

attack_all(A,Of,[B|C]), lock <=>
  Of1 is Of + 1, attack(A,Of,B), attack_all(A, Of1, C), lock.
attack_all(X,Y,[]), lock <=> lock.

draw(0,B,C), confused([D|C]), lock <=> C = [], lock.
draw(A,B,C), confused([D|C]), lock <=>
  C = [E|F],
  A>0,
  G is A-1,
  member(E,B),
  draw(G,B,F),
  attack_all(D,1,[E|F]),
  confused([E|F]),
  lock.

lock <=> true.
\end{Verbatim}
\caption{CHR code for confused queens, without encoding of instantiation}
\label{lst:unsafecqueens}
\end{mylisting}

\section{A more refined synthesis}
\label{sec:refinement}
The CHR program synthesized in Section \ref{sec:confused} for confused queens compiles the coroutines correctly for all top level queries $\mathit{cqueens}(x,y)$, where $x$ is an integer and $y$ a free variable.
Unfortunately, there are instances of the abstract atom $\mathit{cqueens}(g_1, a_1)$ which show that the transformation is unsound.
Furthermore, while our synthesis of permutation sort is sound, a different --- but valid --- abstract analysis would also have led to an unsound synthesis.

The lack of soundness can be demonstrated as follows:
for queries such as $\mathit{cqueens}(4,[X,Y,Z,T])$, where we partially instantiate the output, the CHR program ends with a run-time error.
The reason is that we rely on the CHR matching mechanism to identify whether a constraint $\mathit{attack\_all}/3$ has a sufficiently instantiated third argument to activate rule (5).
If the output argument in $\mathit{cqueens}$ is a variable, then a sufficiently instantiated third argument will have a ground value for its first list member.
But if the output argument is given the required list structure from the start, then CHR is unable to distinguish between $\mathit{attack\_all}$ atoms that have a value ready for use, and those that have the structure, but not the value of the first member.

A trivial way to solve this is to add a guard condition $\mathit{ground}(B)$ to rule (5).
This will provide a correct synthesis for all intended queries.
However, this defeats the purpose of the transformation, as we are then explicitly encoding a kind of ``delay'' in the CHR program, while the aim is to eliminate the delays or freezes from the Prolog program.
Furthermore, this approach prevents portability to CHR implementations not based on Prolog.

Fortunately, the abstract analysis provides us with the knowledge of when an $\mathit{attack\_all/3}$ atom's first list element will be instantiated.
This information can be encoded into the constraints, so that it is available to the matching mechanism.
In the conjunctions in the analysis, there are two $\mathit{multi}$-constructs.
One with $\mathit{attack\_all/3}$ atoms that are sufficiently instantiated, the other with such atoms that are not yet sufficiently instantiated.
In our synthesis, we can take advantage of this information and add guarantees about the instantiation level of the final argument to each $\mathit{attack\_all}/3$, making it an $\mathit{attack\_all/4}$ constraint.
More generally, for every CHR constraint (except for the $\mathit{lock}/0$), we add an additional list argument.
For every element in the original list of arguments, we add a list element representing the guaranteed instantiation of that element to the new argument.
For instance, if an $\mathit{attack\_all}$ constraint is sufficiently instantiated for rewriting, its added fourth argument will be $[g,g,[g|a]]$.

This mainly affects the rules (4) and (5), although it also requires trivial modifications in other rules, as the arity of coroutining constraints increases everywhere and as the instantiation argument must be kept consistent with the abstract analysis.
In rule (5), the newly generated $\mathit{attack\_all}$ is not sufficiently instantiated.
So the rule becomes:

\begin{verbatim}
lock, attack_all(D,E,[F|C],[g,g,[g|a]]) <=>
  H is E + 1, attack(D,E,F), attack_all(D,E,C,[g,g,a]), lock.
\end{verbatim}

In rule (4) the newly generated $\mathit{attack\_all}$ is sufficiently instantiated.
So the rule becomes:

\begin{verbatim}
draw(A,B,C), confused([D|C],[g|a]), lock <=> C = [E|F], A>0, G is A-1,
  member(E,B), draw(G,B,F,[g,g,a]), attack_all(D,1,[E|F],[g,g,[g|a]]),
  confused([E|F],[g|a]), lock.
\end{verbatim}

However, in the abstract analysis we see that in the branches corresponding to this rule, all $\mathit{attack\_all}/3$ atoms that were not sufficiently instantiated become sufficiently instantiated.
If a single specific constraint was modified, it could be rewritten directly.
Due to the use of $multi$, however, the instantiation information cannot be updated for all affected constraints directly.
Therefore, we introduce a new constraint, $\mathit{rename}/0$, in the body of the rule, resulting in:

\begin{verbatim}
draw(A,B,C), confused([D|C],[g|a]), lock <=> C = [E|F], A>0, G is A-1,
  member(E,B), draw(G,B,F,[g,g,a]), rename, attack_all(D,1,[E|F],[g,g,[g|a]]),
  confused([E|F],[g|a]), lock.
\end{verbatim}

In addition, we add the following rules for $\mathit{rename/0}$:

\begin{verbatim}
lock, rename, attack_all(A,B,C,[g,g,a]) <=>
  attack_all(A,B,C,[g,g,[g|a]]), rename,lock.
lock, rename <=> lock.
\end{verbatim}

This will transform all $\mathit{attack\_all/4}$ constraints labelled as insufficiently instantiated into constraints labelled as sufficiently instantiated.
The resulting program is in given in Listing \ref{lst:cqueenswithgroundness}.

\begin{mylisting}[p]
\noindent
\begin{Verbatim}
:- use_module(library(chr)).
:- chr_constraint cqueens/3, draw/4, confused/2, attack_all/4.
:- chr_constraint lock/0, rename/0.
genlist(N,L) :- N >= 1, genlist_acc(N,[],L).
genlist_acc(N,Acc,L) :- N > 1, Nmin is N-1, genlist_acc(Nmin,[N|Acc],L).
genlist_acc(N,Acc,[1|Acc]) :- N is 1. 

attack(A,_,A).
attack(A,Offset,B) :- Diff is A - B, abs(Diff,Offset).

rename, attack_all(A,B,C,[g,g,a]), lock <=>
  attack_all(A,B,C,[g,g,[g|a]]), rename, lock.
rename, lock <=> lock.

cqueens(0,B,[g,a]) <=> genlist(0,C), B = [].
cqueens(A,B,[g,a]) <=>
  B = [E|F],
  genlist(A,C),
  A > 0,
  D is A-1,
  member(E,C),
  draw(D,C,F,[g,g,a]),
  confused([E|F],[g|a]),
  lock.

attack_all(A,Off,[B|C],[g,g,[g|a]]), lock <=>
  Off1 is Off + 1, attack(A,Off,B), attack_all(A,Off1,C,[g,g,a]), lock.
attack_all(X,Y,[],_), lock <=> lock.

draw(0,B,C,[g,g,a]), confused([D|C],[g|a]), lock <=> C = [], lock.
draw(A,B,C,[g,g,a]), confused([D|C],[g|a]), lock <=>
  C = [E|F],
  A>0,
  G is A-1,
  member(E,B),
  draw(G,B,F,[g,g,a]),
  rename,
  attack_all(D,1,[E|F],[g,g,[g|a]]),
  confused([E|F],[g|a]),
  lock.

lock <=> true.
\end{Verbatim}
\caption{CHR code for confused queens, with guaranteed levels of instantiation}
\label{lst:cqueenswithgroundness}
\end{mylisting}

This program correctly compiles the coroutine for any query in the concretization of $\mathit{cqueens}(g_1,a_1)$.
More generally, the technique illustrated above can be proven to be sound and we have successfully applied it to several well-known problems, notably the standard $N$-queens problem, the graph coloring problem, the sieve of Eratosthenes and the sameleaves problem.
The syntheses of these programs contain only simplification rules, which is a desirable property for the application of ranking-based techniques \cite{DBLP:conf/compulog/Fruhwirth99} \cite{DBLP:conf/kr/Fruhwirth02} for analysis of CHR programs.

\section{Conclusions}
We have presented a new application of Compiling Control, which allows coroutining logic programs to be compiled to concise CHR programs with a simple structure.
One could argue that a Prolog program with delays or $\mathit{freeze}$ predicates is also a concise representation.
However, the CHR rules explicitly describe the procedural transitions, which are much harder to see in a Prolog program with delays.
Due to the straightforward implementation of these transitions using simplification rules, the synthesized CHR programs provide a suitable target for further program analysis based on tight rankings \cite{DBLP:conf/compulog/Fruhwirth99}.
Our eventual goal is to apply techniques such as termination analysis \cite{DBLP:conf/compulog/Fruhwirth99} and automatic complexity analysis \cite{DBLP:conf/kr/Fruhwirth02} to the resulting programs.

The resulting CHR programs are slower than their Prolog counterparts --- benchmarks are provided in appendix --- but they are far more readable:
They express only the changes that take place between two states in $\mathcal{A}$, they do not require conjunctions to be renamed using meta-predicates, and they do not require a concrete representation of the $\mathit{multi}$ abstraction.

As a final advantage, our syntheses can be ported to CHR implementations not based on Prolog, such as JCHR or HaskellCHR, provided counterparts to built-in constraints are available.

\nocite{nysccchrappendices2017}

\clearpage
\bibliographystyle{eptcs}
\bibliography{cctochrBibliography}

\newpage
\appendix
\section{Remaining syntheses}
\begin{mylisting}
\noindent
\begin{Verbatim}[frame=none,numbers=left,framesep=3mm]
:- use_module(library(chr)).
:- chr_constraint permsort/3.
:- chr_constraint perm/3.
:- chr_constraint ord/2.
permsort([],X,[g,a]) <=> X = [].
permsort([A|B],Y,[g,a]) <=>
  Y = [C|D],
  select(C,[A|B],E),
  perm(E,D,[g,a]),
  ord([C|D],[g|a]).
perm([],D,[g,a]), ord([C|D],[g|a]) <=> D = [].
perm([F|G],D,[g,a]), ord([C|D],[g|a]) <=>
  D = [H|I],
  select(H,[F|G],J),
  C =< H,
  perm(J,I,[g,a]),
  ord([H|I],[g|a]).
\end{Verbatim}
\caption{CHR code for permutation sort, with guaranteed levels of instantiation}
\label{lst:safepermsort}
\end{mylisting}

\begin{mylisting}
\noindent
\begin{Verbatim}[frame=none,numbers=left,framesep=3mm]
genlist(N,L) :- N >= 1, genlist_acc(N,[],L).
genlist_acc(N,Acc,L) :- N > 1, Nmin is N-1, genlist_acc(Nmin,[N|Acc],L).
genlist_acc(N,Acc,[1|Acc]) :- N is 1.

attack(A,_,A).
attack(A,Offset,B) :-
  Diff is A - B,
  abs(Diff,Offset).

queens(0,[]).
queens(A,[D|E]) :-
  genlist(A,C),
  A > 0,
  F is A - 1,
  member(D,C),
  a(draw(F,C,E),confused([D|E])).

a(draw(0,_,[]),confused([_])).
a(draw(A,B,[E|F]),confused([D,E|F])) :-
  A >= 0,
  G is A - 1,
  member(E,B),
  b(draw(G,B,F),
    multi([attack_all(D,1,[E|F])]),
    confused([E|F])).

b(draw(A,B,C),multi([attack_all(D,E,[F|C])]),confused([F|C])) :-
  H is E + 1,
  attack(D,E,F),
  c(draw(A,B,C),
    multi([attack_all(D,H,C)]),
    confused([F|C])).
b(draw(A,B,C),multi([attack_all(D,E,[F|C]),
  attack_all(H,I,[F|C])|J]),confused([F|C])) :-
  K is E + 1,
  attack(D,E,F),
  d(draw(A,B,C),multi([attack_all(D,K,C)]),
    multi([attack_all(H,I,[F|C])|J]),confused([F|C])).
\end{Verbatim}
\caption{Prolog synthesis of confused queens, part 1}
\label{lst:prologcqueens1}
\end{mylisting}

\begin{mylisting}
\noindent
\begin{Verbatim}[frame=none,numbers=left,framesep=3mm,firstnumber=39]
d(draw(A,B,C),multi([attack_all(D,E,C)|F]),
  multi([attack_all(G,H,[I|C])]),confused([I|C])) :-
  K is H + 1,
  attack(G,H,I),
  append([attack_all(D,E,C)|F],[attack_all(G,K,C)],
         Appended),
  c(draw(A,B,C),multi(Appended),confused([I|C])).
d(draw(A,B,C),multi([attack_all(D,E,C)|F]),
  multi([attack_all(G,H,[I|C]),attack_all(K,L,[I|C])|M]),
  confused([I|C])) :- 
  N is H + 1,
  attack(G,H,I),
  append([attack_all(D,E,C)|F],[attack_all(G,N,C)],
         Appended),
  d(draw(A,B,C),multi(Appended),
    multi([attack_all(K,L,[I|C])|M]),confused([I|C])).

c(draw(0,B,[]),multi([attack_all(D,E,[])|F]),confused([G])) :-
  e(multi([attack_all(D,E,[])|F]),confused([G])).
c(draw(A,B,[H|I]),multi([attack_all(D,E,[H|I])|F]),confused([G,H|I])) :-
  A > 0,
  J is A - 1,
  member(H,B),
  append([attack_all(D,E,[H|I])|F],
         [attack_all(G,1,[H|I])],Appended),
  b(draw(J,B,I),multi(Appended),confused([H|I])).

e(multi([attack_all(A,B,[])]),confused([Z])).
e(multi([attack_all(A,B,[]),
  attack_all(C,D,[])|E]),confused([Z])) :-
  e(multi([attack_all(C,D,[])|E]),confused([Z])).
\end{Verbatim}
\caption{Prolog synthesis of confused queens, part 2}
\label{lst:prologcqueens2}
\end{mylisting}

Listing \ref{lst:safepermsort} shows alternative code for permutation sort, as it would be generated by a fully automated system.
Listings \ref{lst:prologcqueens1} and \ref{lst:prologcqueens2} show a Prolog synthesis of the confused queens problem.

\section{Performance}
Table \ref{table:permsort} shows a performance comparison between the (SWI-)Prolog synthesis and the CHR synthesis of permutation sort.
For each list size from 1 to 20, we generated 10 lists of random elements between 1 and 10 and computed the average time required to generate all solutions.
The fifth and sixth columns shows the relative increase in the number of inferences from the problem size on the next line with respect to the problem size on the current line.
Larger problem sizes could not be analyzed on commodity hardware, but the data strongly suggest both syntheses have the same (exponential) asymptotic complexity, with the CHR version being slower by a factor of less than 4.
A slowdown is to be expected, as CHR uses Prolog as a host language.

Table \ref{table:cqueens} shows a performance comparison between the syntheses of confused queens.
Here, there is no need to determine the average number of inferences, as the solutions depend solely on the number of queens.
The data suggest both syntheses have the same (exponential) asymptotic complexity, though the Prolog ratio moves towards the asymptotic ratio slightly more quickly.
To illustrate this, we have added a column, rratio, which shows how the increase over the previous CHR reading relates to that over the previous Prolog reading.
This number seems to be moving towards $1.00$.
Because it goes down slowly, however, the CHR synthesis has a much larger constant factor.

\begin{table}[]
\centering
\begin{tabular}{llllll}
list size & Inferences (Prolog) & inferences(CHR)  & CHR / Prolog & ratio Prolog & ratio CHR \\
1         & 21.30               & 57.50            & 2.70         & 1.13         & 2.75      \\
2         & 24.00               & 158.00           & 6.58         & 2.11         & 2.66      \\
3         & 50.60               & 420.80           & 8.32         & 2.18         & 2.11      \\
4         & 110.20              & 889.80           & 8.07         & 2.42         & 2.28      \\
5         & 266.20              & 2,032.30         & 7.63         & 2.45         & 2.27      \\
6         & 652.00              & 4,604.70         & 7.06         & 2.57         & 2.40      \\
7         & 1,676.60            & 11,032.40        & 6.58         & 2.45         & 2.28      \\
8         & 4,112.40            & 25,124.80        & 6.11         & 3.01         & 2.91      \\
9         & 12,380.20           & 73,165.20        & 5.91         & 2.85         & 2.74      \\
10        & 35,319.60           & 200,775.20       & 5.68         & 3.94         & 4.02      \\
11        & 139,226.20          & 806,299.20       & 5.79         & 1.62         & 1.42      \\
12        & 225,767.60          & 1,145,682.60     & 5.07         & 2.74         & 2.61      \\
13        & 618,544.20          & 2,985,476.80     & 4.83         & 3.33         & 3.31      \\
14        & 2,057,996.00        & 9,875,311.00     & 4.80         & 4.77         & 5.05      \\
15        & 9,818,654.20        & 49,859,217.20    & 5.08         & 2.75         & 2.55      \\
16        & 27,045,593.20       & 127,279,040.20   & 4.71         & 1.92         & 1.79      \\
17        & 51,887,553.80       & 228,101,140.80   & 4.40         & 3.79         & 3.73      \\
18        & 196,899,172.00      & 850,395,879.00   & 4.32         & 1.26         & 1.17      \\
19        & 247,625,162.20      & 992,932,189.20   & 4.01         & 4.08         & 4.02      \\
20        & 1,010,709,846.00    & 3,996,547,817.00 & 3.95         &              &          
\end{tabular}
\caption{Performance comparison for permutation sort}
\label{table:permsort}
\end{table}

\begin{table}[]
\centering
\begin{tabular}{lllllll}
Queens & Inferences (Prolog) & Inferences (CHR) & CHR/Prolog & ratio Prolog & ratio CHR & rratio \\
5      & 15705               & 55057            & 3.51       & 1.80         & 12.41     & 6.90      \\
10     & 28261               & 683233           & 24.18      & 4.00         & 4.79      & 1.20      \\
15     & 113081              & 3275297          & 28.96      & 2.75         & 3.17      & 1.16      \\
20     & 310581              & 10390703         & 33.46      & 2.22         & 2.51      & 1.13      \\
25     & 690251              & 26037492         & 37.72      & 1.94         & 2.15      & 1.11      \\
30     & 1337851             & 55953873         & 41.82      & 1.76         & 1.93      & 1.10      \\
35     & 2355371             & 107889137        & 45.81      & 1.64         & 1.78      & 1.08      \\
40     & 3861071             & 191885243        & 49.70      & 1.55         & 1.67      & 1.08      \\
45     & 5989441             & 320557732        & 53.52      & 1.48         & 1.59      & 1.07      \\
50     & 8891241             & 509377313        & 57.29      & 1.43         & 1.53      & 1.07      \\
55     & 12733461            & 776950777        & 61.02      & 1.39         & 1.47      & 1.06      \\
60     & 17699361            & 1145302583       & 64.71      & 1.36         & 1.43      & 1.06      \\
65     & 23988431            & 1640155772       & 68.37      & 1.33         & 1.40      & 1.05      \\
70     & 31816431            & 2291213553       & 72.01      & 1.30         & 1.37      & 1.05      \\
75     & 41415351            & 3132440217       & 75.63      & 1.28         & 1.34      & 1.05      \\
80     & 53033451            & 4202342723       & 79.24      & 1.26         & 1.32      & 1.05      \\
85     & 66935221            & 5544251612       & 82.83      & 1.25         & 1.30      & 1.04      \\
90     & 83401421            & 7206602593       & 86.41      & 1.23         & 1.28      & 1.04      \\
95     & 102729041           & 9243217457       & 89.98      & 1.22         & 1.27      & 1.04      \\
100    & 125231341           & 11713585663      & 93.54      & 1.21         & 1.25      & 1.04      \\
105    & 151237811           & 14683145252      & 97.09      & 1.20         & 1.24      & 1.04      \\
110    & 181094211           & 18223564433      & 100.63     & 1.19         & 1.23      & 1.04      \\
115    & 215162531           & 22413022497      & 104.17     & 1.18         & 1.22      & 1.03      \\
120    & 253821031           & 27336491403      & 107.70     & 1.17         & 1.21      & 1.03      \\
125    & 297464201           & 33086016692      & 111.23     & 1.16         & 1.20      & 1.03      \\
130    & 346502801           & 39760999073      & 114.75     & 1.16         & 1.19      & 1.03      \\
135    & 401363821           & 47468475337      & 118.27     & 1.15         & 1.19      & 1.03      \\
140    & 462490521           & 56323399943      & 121.78     & 1.15         & 1.18      & 1.03      \\
145    & 530342391           & 66448925932      & 125.29     & 1.14         & 1.17      & 1.03      \\
150    & 605408035           & 77976686518      & 128.80     &              &           &        
\end{tabular}
\caption{Performance comparison for confused queens}
\label{table:cqueens}
\end{table}

\section{Abstract analysis of confused queens}

\begin{figure}[h!]
  \centering
\makebox[\textwidth][c]{
  \includegraphics[scale=0.9]{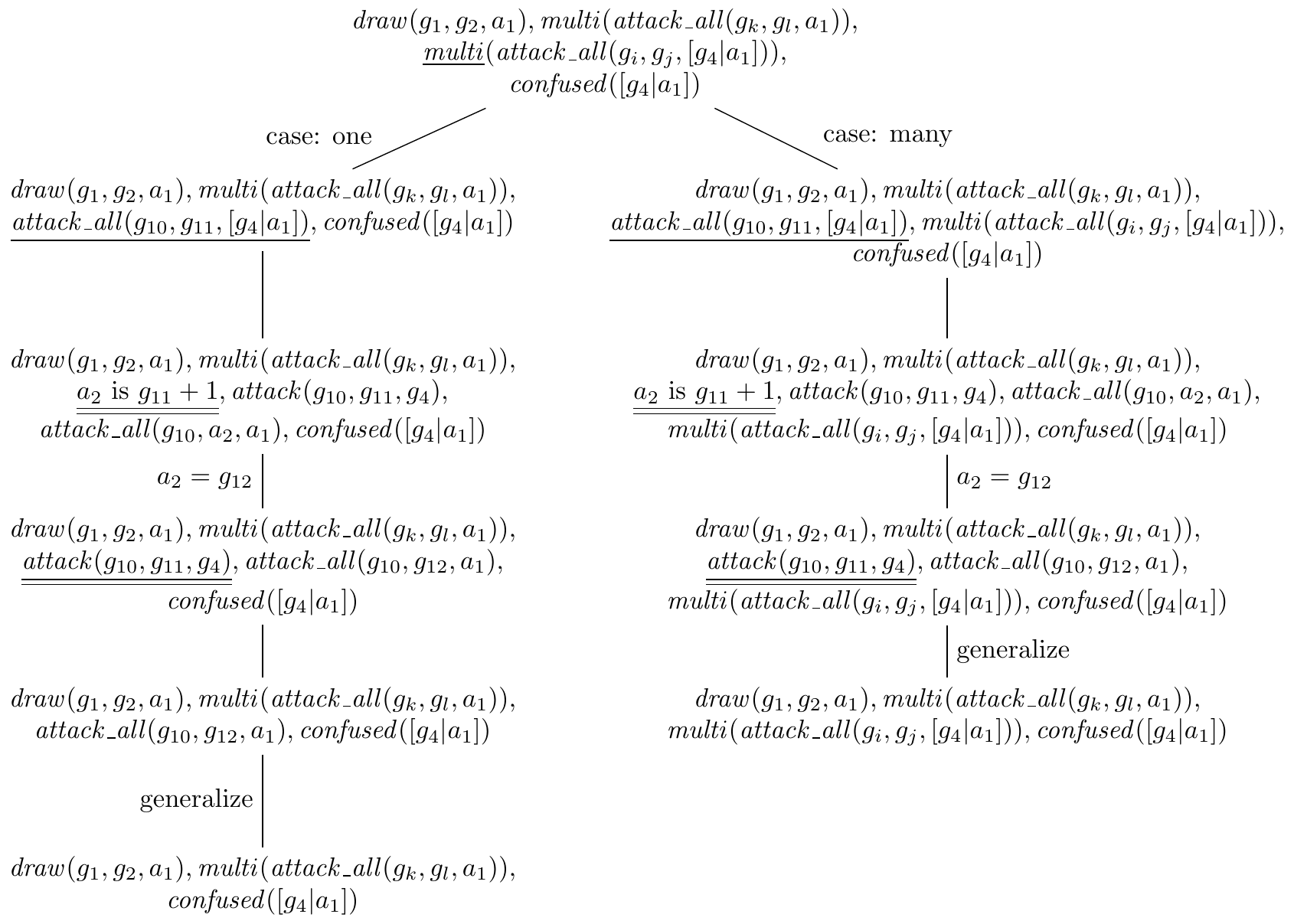}
}
  \caption{Fourth abstract tree in the analysis of $\mathit{cqueens}(g_1,a_1)$.}
  \label{fig:abstracttreecqueens4a}
\end{figure}

\begin{figure}[h!]
  \centering
\makebox[\textwidth][c]{
  \includegraphics[scale=0.9]{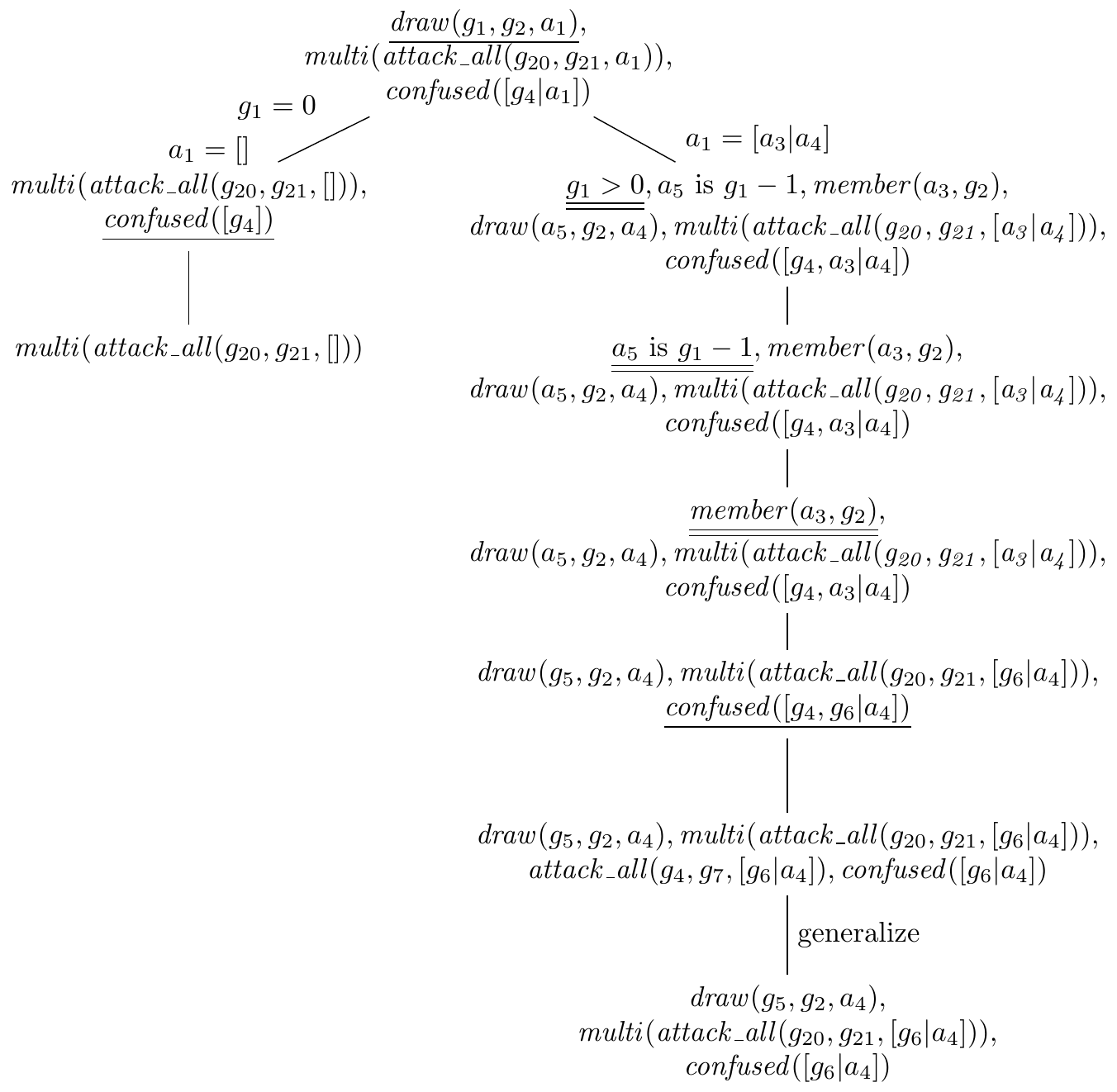}
}
  \caption{Fifth abstract tree in the analysis of $\mathit{cqueens}(g_1,a_1)$.}
  \label{fig:abstracttreecqueens5a}
\end{figure}

\begin{figure}[h!]
  \centering
\makebox[\textwidth][c]{
  \includegraphics[scale=0.9]{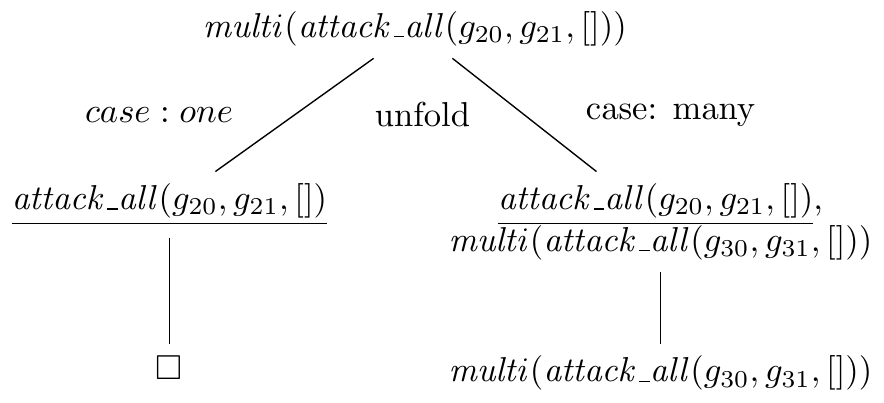}
}
  \caption{Sixth abstract tree in the analysis of $\mathit{cqueens}(g_1,a_1)$.}
  \label{fig:abstracttreecqueens6a}
\end{figure}

Figures \ref{fig:abstracttreecqueens4a}, \ref{fig:abstracttreecqueens5a} and \ref{fig:abstracttreecqueens6a} show the remaining abstract analysis trees for confused queens.

\newpage

\section{Concrete analysis of confused queens}
\begin{figure}[h!]
  \centering
\makebox[\textwidth][c]{
  \includegraphics[scale=0.9]{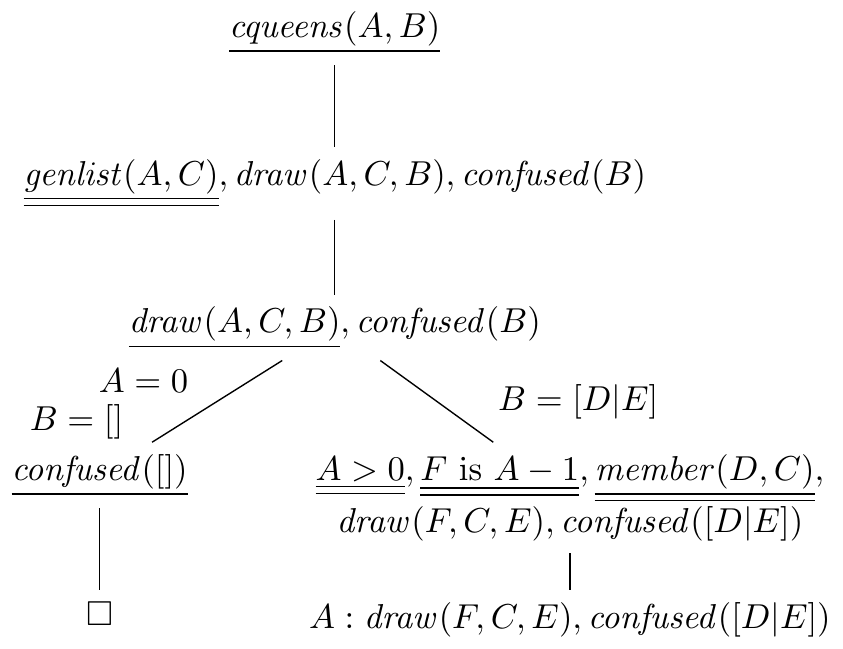}
}
  \caption{Concrete tree corresponding to Figure \ref{fig:abstracttreecqueens1}.}
  \label{fig:concretetreecqueens1a}
\end{figure}

\begin{figure}[h!]
  \centering
\makebox[\textwidth][c]{
  \includegraphics[scale=0.9]{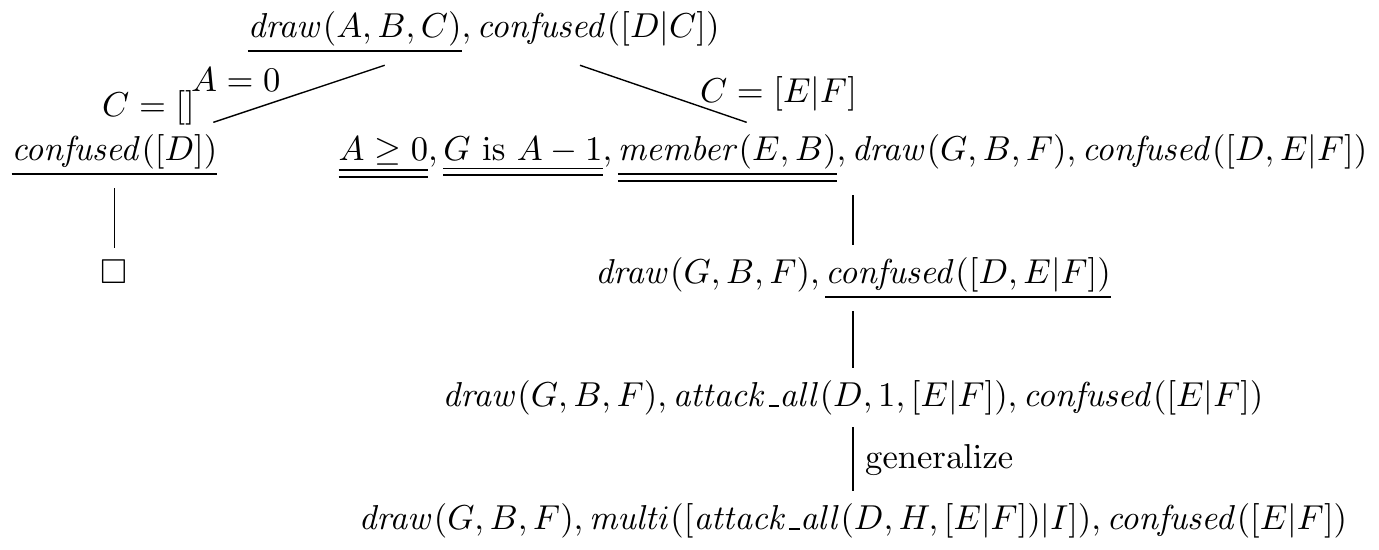}
}
  \caption{Concrete tree corresponding to Figure \ref{fig:abstracttreecqueens2}.}
  \label{fig:concretetreecqueens2a}
\end{figure}

\begin{figure}[h!]
  \centering
\makebox[\textwidth][c]{
  \includegraphics[scale=0.9]{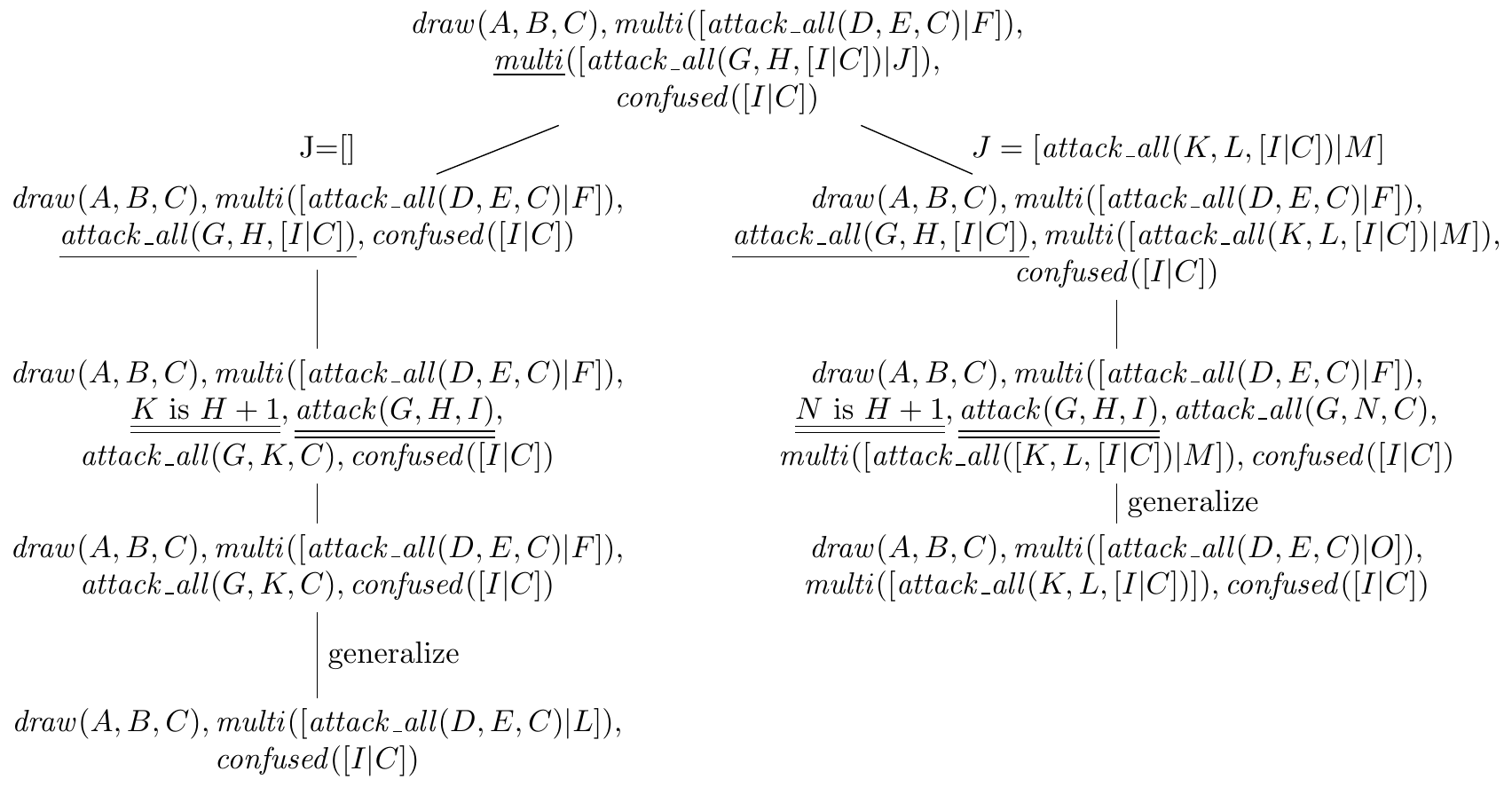}
}
  \caption{Concrete tree corresponding to Figure \ref{fig:abstracttreecqueens4a}.}
  \label{fig:concretetreecqueens4a}
\end{figure}

\begin{figure}[h!]
  \centering
\makebox[\textwidth][c]{
  \includegraphics[scale=0.9]{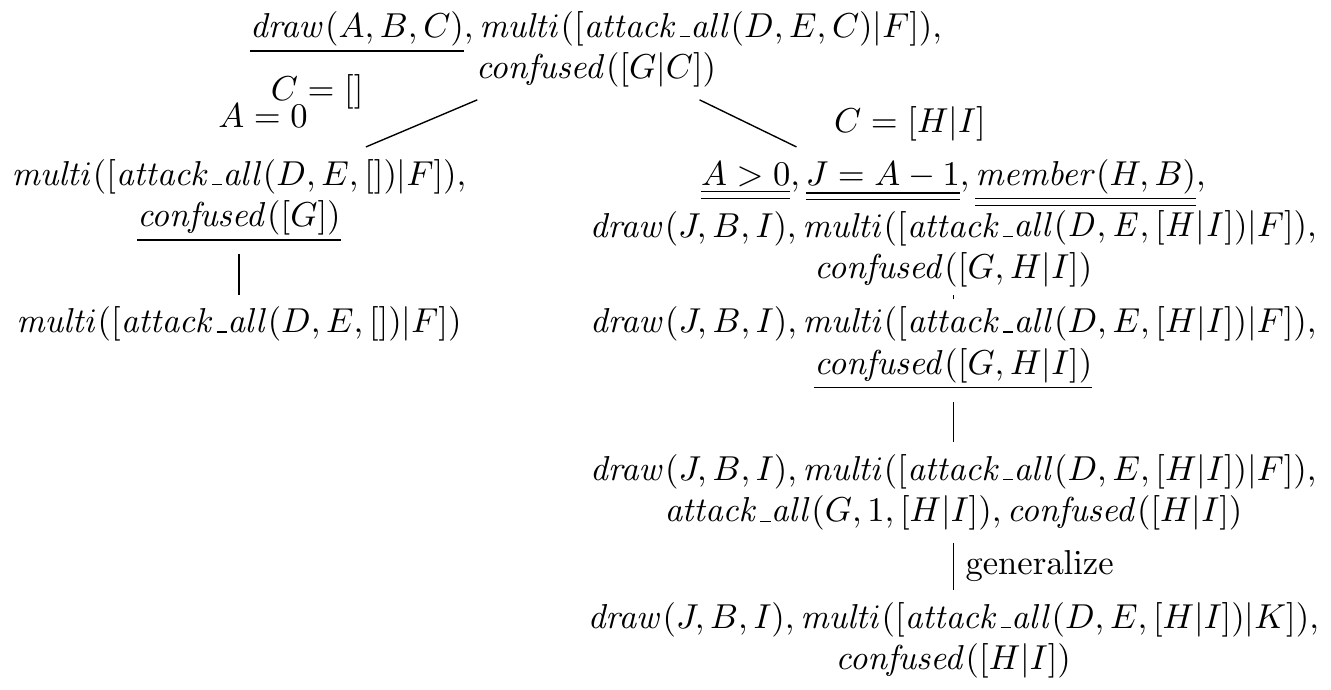}
}
  \caption{Concrete tree corresponding to Figure \ref{fig:abstracttreecqueens5a}.}
  \label{fig:concretetreecqueens5a}
\end{figure}

\begin{figure}[h!]
  \centering
\makebox[\textwidth][c]{
  \includegraphics[scale=0.9]{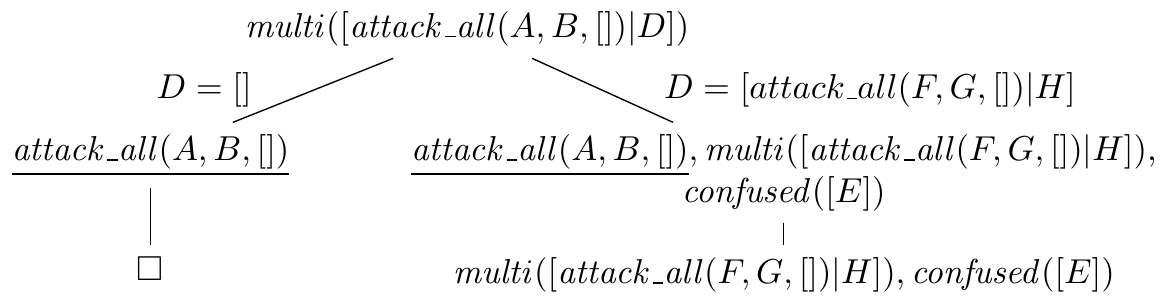}
}
  \caption{Concrete tree corresponding to Figure \ref{fig:abstracttreecqueens6a}.}
  \label{fig:concretetreecqueens6a}
\end{figure}

Figures \ref{fig:concretetreecqueens1a} through \ref{fig:concretetreecqueens6a} show the concrete trees for the confused queens problem.

\begin{figure}[h!]
  \centering
\makebox[\textwidth][c]{
  \includegraphics[scale=0.9]{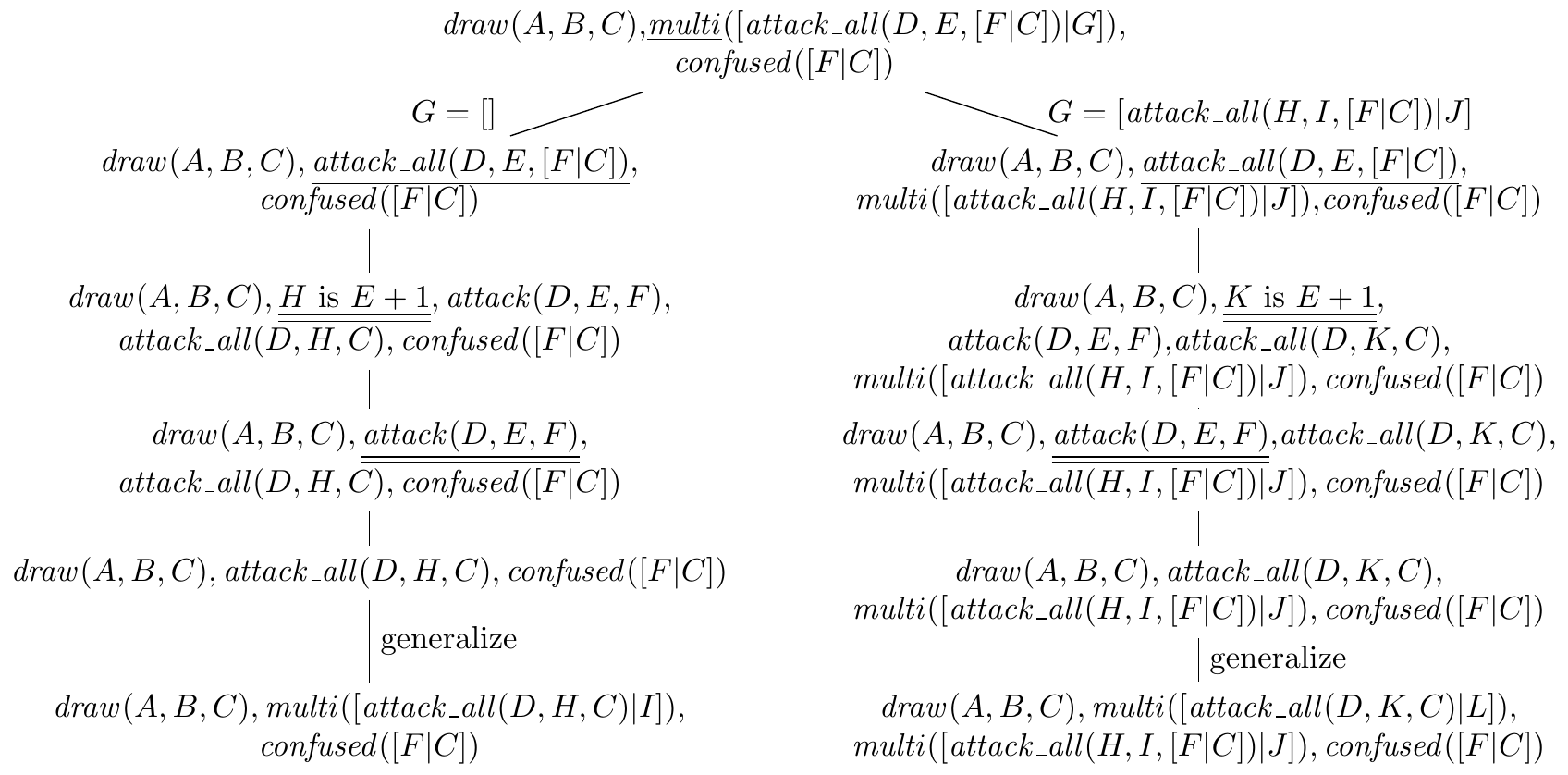}
}
  \caption{Concrete tree corresponding to Figure \ref{fig:abstracttreecqueens3}.}
  \label{fig:concretetreecqueens3a}
\end{figure}

\newpage
% \lstinputlisting[caption=unsafe CHR synthesis of confused queens,language={},float=h]{code/chrconfusedqueens.pl}
% \lstinputlisting[caption=safe CHR synthesis of confused queens,language={},float=h]{code/chrconfusedqueenswithid.pl}
\FloatBarrier % prevents listing from being mixed with trees
% \lstinputlisting[caption=Prolog synthesis of confused queens,language={}]{code/prologsynthesisconfusedqueens.pl}

\end{document}